\documentclass{aa}  

\usepackage{graphicx} 
\usepackage{linenoaa}
\usepackage{txfonts}
\begin{document}

   \title{Chemical differences among collapsing low-mass protostellar cores}

   \subtitle{}

   \author{Jingfei Sun\inst{1}, Xiaohu Li\inst{1,2}, Fujun Du\inst{3,4}, Yao Wang\inst{3}, Juan Tuo\inst{1,5}, Yanan Feng\inst{1,5}
          }

   \institute{
         Xinjiang Astronomical Observatory, Chinese Academy of Sciences, 150 Science 1-Street, Urumqi 830011, P.R.China\\
         \email{sunjingfei@xao.ac.cn; xiaohu.li@xao.ac.cn}
         \and
         Key Laboratory of Radio Astronomy, Chinese Academy of Sciences, Urumqi 830011, P.R.China
         \and 
         Purple Mountain Observatory and Key Laboratory of Radio Astronomy, Chinese Academy of Sciences, Nanjing 210023, P.R.China
         \and
         School of Astronomy and Space Science, University of Science and Technology of China, Hefei 230026, P.R.China
         \and
         University of Chinese Academy of Sciences, P.R.China
             }

   \date{Received XXX; accepted XXX}

 
  \abstract
   {Organic features lead to two distinct types of Class 0/I low-mass protostars: hot corino sources, exhibiting abundant saturated complex organic molecules (COMs), and warm carbon-chain chemistry (WCCC) sources, exhibiting abundant unsaturated carbon-chain molecules.
   Some observations suggest that the chemical variations between WCCC sources and hot corino sources are associated with local environments, as well as the luminosity of protostars.}
   {We aim to investigate the physical conditions that significantly affect WCCC and hot corino chemistry, and to reproduce the chemical characteristics of prototypical WCCC sources and hybrid sources, where both carbon-chain molecules and COMs are abundant.
   }
   {We conducted gas-grain chemical simulation in collapsing protostellar cores, adopting some typical physical parameters for the fiducial model.
   By changing values of some physical parameters, including the visual extinction of ambient clouds ($A_{\rm V}^{\rm amb}$), the cosmic-ray ionization rate ($\zeta$), the maximum temperature during the warm-up phase ($T_{\rm max}$), and the contraction timescale of protostars ($t_{\rm cont}$), we studied the dependence of WCCC and hot corino chemistry on these physical parameters.
   Subsequently, we ran a model with different physical parameters to reproduce scarce COMs in prototypical WCCC sources.
   }
   {The fiducial model predicts abundant carbon-chain molecules and COMs, and reproduces WCCC and hot corino chemistry in the hybrid source L483.
   This suggests that WCCC and hot corino chemistry can coexist in some hybrid sources.
   Ultraviolet (UV) photons and cosmic rays can boost WCCC features by accelerating the dissociation of CO and CH$_4$ molecules.
   On the other hand, UV photons can weaken the hot corino chemistry by photodissociation reactions, while the dependence of hot corino chemistry on cosmic rays is relatively complex.
   The $T_{\rm max}$ does not affect WCCC features, while it can influence hot corino chemistry by changing the effective duration of two-body surface reactions for most COMs.
   The long $t_{\rm cont}$ can boost WCCC and hot corino chemistry, by prolonging the effective duration of WCCC reactions in the gas phase and surface formation reactions for COMs, respectively.
   The scarcity of COMs in prototypical WCCC sources can be explained by insufficient dust temperature in the inner envelopes to activate hot corino chemistry.
   Meanwhile, the High $\zeta$ and the long $t_{\rm cont}$ favors the explanation for scarce COMs in these sources. 
   }
   {The chemical differences between WCCC sources and hot corino sources can be attributed to the variations in local environments, such as $A_{\rm V}^{\rm amb}$, $\zeta$, as well as the protostellar property $t_{\rm cont}$.
   }
   \keywords{Astrochemistry -- Stars: protostars -- ISM: abundances -- ISM: evolution -- ISM: clouds -- ISM: molecules
               }

   \authorrunning{J.F. Sun et al.}            
   \titlerunning{Chemical differences among low-mass protostellar cores}  
   \maketitle
%
\nolinenumbers

\section{Introduction}           
\label{sec_introduction}
Carbon-bearing molecules consisting of six or more atoms are considered as complex organic molecules (COMs, \citealt{Herbst+vanDishoeck+2009,Jorgensen+etal+2020,Ceccarelli+etal+2023}), and some of them suggest a possible relation between interstellar chemistry and the emergence of life on Earth.
COMs have been detected in many objects, including prestellar or starless cores (e.g., \citealt{Jimenez-Serra+etal+2016,Scibelli+etal+2021}), protostellar cores (e.g., \citealt{Chahine+etal+2022}), protostellar disks (e.g., \citealt{Lee+etal+2019}), and protoplanetary disks (e.g., \citealt{Loomis+etal+2018}).
The chemistry in low-mass protostellar cores is a main topic of astronomy, as it determines the interstellar chemical legacy inherited by protoplanetary disks.
During the early phases of low-mass protostellar cores, namely Class 0 and I, organic characteristics lead to two special types of protostars: hot corinos (e.g., \citealt{Cazaux+etal+2003,Ceccarelli+etal+2017}) and warm carbon-chain chemistry sources (WCCC; e.g., \citealt{Sakai+etal+2008,Hirota+etal+2009}). 

Hot corinos are compact sources (with sizes $\leq$ 100 au) with high temperatures ($\geq$ 100 K), showing abundant presence of saturated COMs in hot central regions (e.g, \citealt{vanDishoeck+etal+1995,Ceccarelli+2004,Caux+etal+2011,Taquet+etal+2015,Jorgensen+etal+2016,Lopez-Sepulcre+etal+2017}).
Currently, more than twenty COMs (e.g., CH$_3$OH, CH$_3$CN, CH$_3$OCH$_3$, NH$_2$CHO, etc.) have been detected in hot corinos (\citealt{Imai+etal+2016,Jorgensen+etal+2016,Lopez-Sepulcre+etal+2017,Marcelino+etal+2018a,Bianchi+etal+2019,Belloche+etal+2020,Nazari+etal+2021,Chahine+etal+2022}). 
The first detected hot corino was in IRAS16293-2422 (referred to as IRAS16293; \citealt{Cazaux+etal+2003}), which contains two hot corinos, A and B (\citealt{Kuan+etal+2004,Bottinelli+etal+2004b}).
Later, other hot corinos were found towards NGC1333-IRAS4A (\citealt{Bottinelli+etal+2004a}), NGC1333-IRAS2A (\citealt{Jorgensen+etal+2005}), and NGC1333-IRAS4B (\citealt{Sakai+etal+2006,Bottinelli+etal+2007}), which have similar chemical characteristics (\citealt{Herbst+vanDishoeck+2009}), with abundant saturated COMs and scarce unsaturated carbon-chain molecules.

Warm carbon-chain chemistry sources represent another class of low-mass Class 0/I protostellar sources. 
They are characterized by high excitation conditions, central concentration, and abundant unsaturated carbon-chain molecules such as hydrocarbons (e.g., C$_{n}$H, C$_{n}$H$_2$, $n = 1, 2, \cdots$), cyanopolyynes (HC$_{2n+1}$N, $n = 1, 2, \cdots$) families and so on (\citealt{Sakai+etal+2009}).
In contrast, they exhibit a relatively limited presence of saturated COMs (e.g., \citealt{Sakai+etal+2008,Sakai+etal+2009,Hirota+etal+2009}). 
These sources are also associated with dense and lukewarm circumstellar envelopes and an enhancement of gaseous CO$_2$ (\citealt{Sakai+etal+2009}).
The first confirmed WCCC source is L1527, which contains the low-mass Class 0 or I protostar IRAS 04368+2557 (\citealt{Sakai+etal+2008,Ohashi+etal+1997,Sakai+Yamamoto+2013}).
Later observations confirm additional sources, such as IRAS 15398--3359 in Lupus and IRAS 18148--0440 in L483, as WCCC sources or candidates, which exhibit abundant carbon-chain molecules (\citealt{Hirota+etal+2009,Sakai+etal+2009,Higuchi+etal+2018}).  
These carbon-chain molecules in warm, extended, dense regions are formed through the evaporation of CH$_4$ ice from dust grains (\citealt{Sakai+etal+2009,Sakai+Yamamoto+2013}).

Distinctive chemical compositions are evident in these prototypical sources: the hot corino IRAS 16293 exhibits a dearth of carbon-chain molecules, whereas the WCCC source L1527 does not display any detectable COMs (\citealt{Sakai+Yamamoto+2013}).
To date, two major reasons have been proposed to explain their chemical variations: different timescale of prestellar cores (\citealt{Sakai+etal+2008,Sakai+Yamamoto+2013}) disproved by the model of \cite{Aikawa+etal+2020}, and different illumination by the UV radiation field (\citealt{Spezzano+etal+2016,Kalvans+2021}).
Some observations indicate that WCCC sources are found in cloud peripheries, while typical hot-corino-like sources are absent in these regions, and instead are located inside dense filamentary clouds (\citealt{Lefloch+etal+2018,Higuchi+etal+2018}).
These observations suggest that carbon-chain molecules are more abundant in protostars with lower visual extinction of ambient clouds, which is consistent with the chemical simulations reported by \cite{Aikawa+etal+2020}.
\cite{Bouvier+etal+2022} found that the OMC-2/3 filament appears to have a relatively lower detection rate of hot corinos (i.e., $\sim$23\%), in comparison with the Perseus low-mass star-forming region, where hot corinos were detected in approximately 60\% of sources by the Perseus ALMA Chemistry Survey (\citealt{Yang+etal+2021}).
Studying in detail the influence of UV photons illumination and cosmic-ray irradiation on WCCC sources, \cite{Kalvans+2021} found that the abundances of carbon-chain molecules can be influenced by cosmic ray and UV irradiation, while the abundances of COMs have no clear correlation with radiation.

Some hybrid sources are also identified by interferometer observations (e.g., \citealt{Graninger+etal+2016,Imai+etal+2016,Oya+etal+2017,Jacobsen+etal+2019,Bouvier+etal+2020}), in which the emissions of carbon-chain species are from extended envelope ($r\sim$ a few $10^2-10^3$ AU), and emissions of COMs are centrally concentrated ($r\lesssim 100$ AU). 
Additionally, by analyzing the C${_2}$H/CH$_3$OH ratio, \cite{Higuchi+etal+2018} discovered that most sources display intermediate chemical features between these two distinct types.
There are only a few theoretical works investigating both WCCC and hot corino chemistry (e.g., \citealt{Aikawa+etal+2020}), but reproducing the scarce saturated COMs in canonical WCCC sources is challenging. 

Dynamics is essential for studying the chemical compositions of protostellar cores.
There are numerous theoretical and numerical simulation approaches to low-mass star formation, such as the semianalytical L-P model (\citealt{Larson+1969,Penston+1969}), self-similar expansion wave collapse model (\citealt{Shu+1977}), complicated 1-dimensional radiation hydrodynamical model (\citealt{Masunaga+Inutsuka+2000}), 2-dimensional hydrodynamical model, and full 3-dimensional radiation magnetohydrodynamics model (\citealt{Tomida+etal+2010}).
On the other hand, numerous works have indicated that carbon-chain molecules and COMs in the protostellar cores are mainly originated from infalling envelopes rather than rotating disks (e.g., \citealt{Myers+etal+1995,Pineda+etal+2012,Imai+etal+2019,Jacobsen+etal+2019,Belloche+etal+2020}). 
Therefore, the expansion wave collapse model can satisfy our research aims to study molecular spatial distributions, without incurring enormous computational costs as radiation hydrodynamical simulations (e.g., \citealt{Aikawa+etal+2008,Aikawa+etal+2020}).

This paper is organized as follows.
In Sect.~\ref{sect_model}, we describe in detail physical and chemical models.
To explore whether hybrid sources are a common occurrence, we reproduce the chemical characteristics of carbon-chain molecules and COMs in the fiducial model in Sect.~\ref{sect_Results}.
Additionally, we compare our results with some observations, including a hybrid source L483.
To find out the potential reasons for chemical diversity between WCCC and hot corino sources, we explore the influence of several physical parameters on carbon-chain molecules and COMs in Sect.~\ref{DependenceOnSomePhysicalParameters}, including the visual extinction of ambient clouds, the cosmic ray ionization rate, the maximum temperature at hot corino phase, and the contraction timescale of the protostar (i.e., warm-up timescale).  
We discuss the reasons for the scarce COMs in the prototypical WCCC sources (e.g., L1527, IRAS16293) and potential effect of some uncertainties on simulated results in Sect.~\ref{sect_Discussion}, followed by a brief conclusion in Sect.~\ref{sect_Conclusions}.

\section{Model}
\label{sect_model}
\subsection{Physical Model}
\label{sect_physical model}
\begin{figure*} 
  \centering
  \includegraphics[width=\hsize, angle=0]{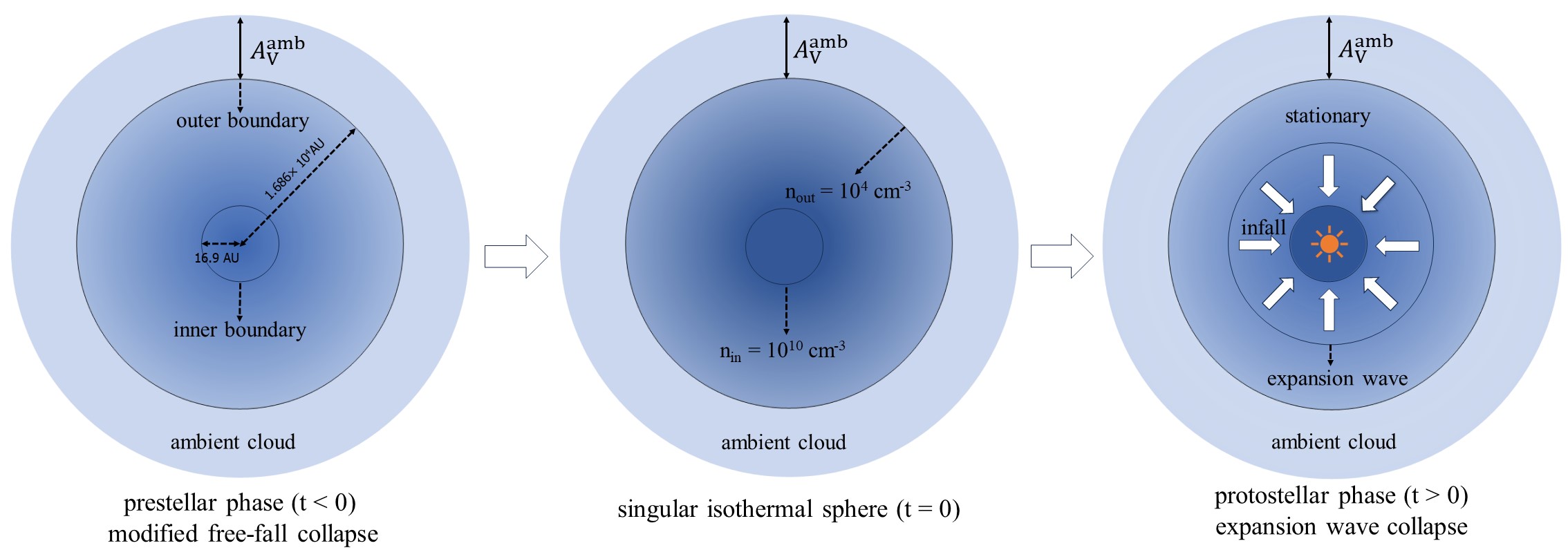}
  \caption{\centering Schematic view of the core model.} 
  \label{SchematicView}
\end{figure*}
In this paper, we adopted a 1-D spherical modified free-fall collapse (\citealt{Spitzer+1978,Rawlings+etal+1992}), followed by an expansion wave collapse (\citealt{Shu+1977}). 
The modified free-fall collapse, with a specific initial number density $n_0$, is depicted by \cite{Rawlings+etal+1992} as follows:
\begin{equation}
  \frac{dn}{dt} = B\left(\frac{n^4}{n_0}\right)^{1/3}\left(24\pi Gm_{\rm H}n_0\left[(\frac{n}{n_0})^{1/3}-1\right]\right)^{1/2}, 
  \label{eq_ModifiedFreeFallDensity}
\end{equation}
where $t$ represents time, $G$ is the gravitational constant, $B$ is the retardation factor, and $m_{\rm H}$ is the mass of a hydrogen atom, respectively.
Initially, the number density at any radius is 10$^4$ cm$^{-3}$, then the modified free-fall collapse would be truncated by a singular isothermal sphere.
The density function of the singular isothermal sphere is given by \cite{Shu+1977}:
\begin{equation}
  \rho(r) = \frac{a^2}{2\pi G}r^{-2}, 
  \label{eq_SingularIsothermalDensity}
\end{equation}
where $a$ represents the sound speed, $G$ is the gravitational constant, and $r$ stands for the radial distance from the core center.
In most cloud cores, magnetic fields and turbulence are very weak relative to gravity (\citealt{Shu+etal+1987}), so we neglected their influence on the sound speed.
When considering a mean molecular weight of 1.36, mainly contributed by hydrogen and helium, and assuming a constant kinetic temperature of 10 K, the sound speed is approximately 0.25 km/s.
Consequently, the outer boundary is fixed at a radius of $1.686{\times}10^{4}$ AU, which corresponds to a number density of $10^{4}$ cm$^{-3}$.
The total mass contained within the outer boundary is about 2.3 M$_\odot$.
Given that we mainly focus on the chemical evolution in the outer envelopes, we fixed the inner boundary at $\sim16.9$ AU (corresponding to a number density of 10$^{10}$ cm$^{-3}$). 

Our physical model consists of two main phases (Fig.~\ref{SchematicView}): the prestellar phase (modified free-fall collapse) and the protostellar phase (expansion wave collapse). 
According to the variation of the temperature, the protostellar phase can be divided into the isothermal collapse phase, warm-up phase, and hot corino phase. 
Their timescales in the fiducial model are presented in Table~\ref{BasicParameters}.
In the modified free-fall collapse, all gas parcels are stationary, but the density at any position increases with time. 
Once the matter density at a given radius reaches the density of the singular isothermal sphere, the density keeps constant until the density profile of the singular isothermal sphere is completed from the outside to the inside at all radii (i.e., from 16.9 to $1.686{\times}10^{4}$ AU). 
Considering the significant effects of magnetic and rotational support at the early stages, we set a moderate retardation factor $B$ equal to 0.7.
The modified free-fall collapse lasts for $7.3{\times}10^5$ yr, during which molecular abundances at the terminal stage are set to the initial molecular abundances of the expansion wave collapse.

After the singular isothermal configure is completed, the cloud core starts to undergo an expansion wave collapse (for a detailed description, see \citealt{Shu+1977}). 
Inside the expansion wave, the matter approaches free-fall with a density distribution of $n \propto r^{-3/2}$.
Outside the expansion wave, the material remains motionless at original radius, with a density distribution of $n \propto r^{-2}$.
We need to follow every gas parcel that undergoes varying physical environments, and calculate the molecular abundances in a Lagrangian coordinate system. 
After the calculation, all parcels are transferred to Eulerian coordinates for each given time step to determine the abundance profiles of molecules with time.
We assume that a gas parcel falls onto the central protostar if the radius of the parcel is less than the inner radius.
 
During the early stages of collapse, cooling by radiation is more efficient than heating by gravitational compression, so the core is nearly isothermal with a kinetic temperature of 10 K (\citealt{Lee+etal+2004}).
The critical density of $10^{11}$ cm$^{-3}$ has been determined by many models and observations, at which point the dust emission becomes opaque (e.g., \citealt{Larson+1969}).
However, as the expansion wave collapse model cannot calculate the isothermal collapse timescale, we adopt an isothermal collapse duration of $2{\times}10^4$ yr artificially (\citealt{Lee+etal+2004}).
When heating overtakes cooling, the temperature in the central regions begins to increase. 
With the gas temperature increasing, the first and second hydrostatic core (i.e., protostar) forms, respectively.
Subsequently the envelopes surrounding the central protostar enter the warm-up phase, the gas and dust temperatures in the infalling envelopes are dominated by the accretion luminosity, disk luminosity, and protostellar luminosity.
We consider the contraction time of a medium-mass protostar of $2{\times}10^5$ yr (\citealt{Bernasconi+Maeder+1996,Viti+Williams+1999}), as the warm-up timescale in the fiducial model.
Afterward, the gas and dust temperatures in infalling envelopes remain constant until the end of chemical evolution (i.e., $t = 5{\times}10^5$ yr, \citealt{Lee+etal+2004}).

Collisions between gas and dust particles are frequent in dense regions ($> 10^4$ cm$^{-3}$), so the gas temperature tends to be coupled with the dust temperature (\citealt{Lee+etal+2004,Bergin+Tafalla+2007}).
For simplicity, we assume that the gas temperature is the same as the dust temperature in all models.
In modified free-fall collapse (i.e., prestellar core), we use a fitting formula to obtain dust temperature that varies with the visual extinction and interstellar radiation fields (\citealt{Hocuk+etal+2017}).
This formula is better suited to fit observations than other formulas (e.g., \citealt{Hollenbach+etal+1991,Zucconi+etal+2001,Garrod+Pauly+2011}).
During the warm-up phase, we use an analytical formula to obtain the dust temperature that varies with time and radius, rather than using dust continuum radiative transfer method. This formula is given by \cite{Rowan-Robinson+1980,Viti+Williams+1999,Garrod+Herbst+2006} as:
\begin{equation}
  T_{\rm dust} = T_0 + (T_{\rm max} - T_0)\left(\frac{t-t_{\rm warm}^0}{t_c}\right)^n(\frac{r}{r_0})^{-0.4},
  \label{eq_DustTemperatureInWarmUpPhase}
\end{equation}
where $T_{\rm dust}$ is the dust temperature, $T_0$ is the initial temperature, 7 K (\citealt{Aikawa+etal+2008,Masunaga+Inutsuka+2000}), $T_{\rm max}$ is the maximum temperature at the end of the warm-up phase (200 K), $t_c = 2{\times}10^5$ yr is the contraction time for intermediate star formation (\citealt{Bernasconi+Maeder+1996,Molinari+etal+2000,Garrod+Herbst+2006}), and $n = 2$ for power-law temperature profiles (\citealt{Viti+etal+2004,Garrod+Herbst+2006}).
After the warm-up phase, the dust temperature at any given radius remains constant until the end of chemical evolution.

At deep positions of the cloud core, cosmic rays undergo attenuation to some extent (\citealt{Padovani+etal+2018}), which dominates molecular chemistry relative to UV photons.
Furthermore, cosmic rays significantly affect WCCC and hot corino chemistry in protostellar cores (see Sect.~\ref{CosmicRay}), so we adopted a fitting formula that varies with the total column density $N_{\rm H}$, rather than treating it as a constant. 
This formula is given by \cite{Padovani+etal+2018} as:
\begin{equation}
  {\rm log}_{10}\zeta = \sum_{k \geq 0}c_k{\rm log}_{10}^{k}N_{\rm H},  
  \label{eq_CosmicRayIonizationRate}
\end{equation}
where $c_k$ are coefficients of the polynomial fit (see model $L$ in Table~F.1 from \citealt{Padovani+etal+2018}).
We assume that the cloud core is exposed to the \cite{Draine+1978} radiation field, which is equal to approximately 1.7 times the \cite{Habing+1968} local interstellar radiation field.
Outside the cloud core, we adopt a high visual extinction of ambient clouds of 3.0 mag (\citealt{Aikawa+etal+2020}).
The relation between the visual extinction $A_{\rm V}$ and the total column density of protons $N_{\rm H}$ is given by \cite{Bohlin+etal+1978} as $A_{\rm V} = 5.34{\times}10^{-22}N_{\rm H}$.
\begin{table}
  \begin{center}
  \caption{Fixed Various Timescales and Other Physical Parameters in the Fiducial Model\label{BasicParameters}}
  \setlength{\tabcolsep}{3pt}
  \begin{tabular}{lc}
  \hline
  \hline\noalign{\smallskip}
  Prestellar timescale & $7.3{\times}10^5$ yr\\
  Isothermal collapse timescale & $2.0{\times}10^4$ yr\\
  Warm-up timescale/Contraction timescale & $2.0{\times}10^5$ yr\\
  Hot corino timescale & $2.8{\times}10^5$ yr\\
  Protostellar timescale & $5.0{\times}10^5$ yr\\
  \hline\noalign{\smallskip}
  Visual extinction of ambient clouds & 3.0 mag\\
  Maximum temperature & 200 K\\
  Unattenuated FUV flux & 1.7 Habing\\
  Dust-to-gas mass ratio & 0.01\\
  Dust albedo & 0.6\\
  Dust grain radius & 0.1 $\mu$m\\
  Dust material density & 3.0 g cm$^{-3}$\\
  Dust site density & $10^{15}$ cm$^{-2}$\\
  Chemical desorption efficiency & 0.03\\
  Ratio of surface diffusion to desorption energies & 0.4\\
  Ratio of bulk diffusion to desorption energies & 0.8\\
  Mean molecular weight & 1.36\\
  \hline\noalign{\smallskip}
  \end{tabular}
\end{center}
\end{table}

\subsection{Chemical Model}
\label{sect_chemical model}
The chemical models in this study are carried out using the astrochemical code Chempl (\citealt{Du+2021}), which supports the three-phase formulation of interstellar gas-grain chemistry following the approach of \cite{Hasegawa+Herbst+1993} and \cite{Garrod+Pauly+2011}. 
In this code, species in the mantle are completely inert, and the transition of species on dust grain surfaces to mantles is achieved through the exposure and coverage (for a detailed description, see \citealt{Du+2021}).
For the gas-phase chemistry, \cite{Du+2021} used the UMIST database for astrochemistry 2012 network (UDFA 12, \citealt{McElroy+etal+2013}). 
Their surface reactions mainly involve the formation of HCHO, CH$_3$OH, H$_2$O, H$_2$O$_2$, and a few additional reactions, which are a combination of selected reactions from some other works (e.g., \citealt{Allen+Robinson+1977,Tielens+Hagen+1982,Hasegawa+etal+1992}).
To meet our research requirements in the hot chemistry, we have implemented several modifications as follows.

Some calculations and experiments have shown that sub-surface processes, such as photodissociation, diffusion, and recombination, can operate in interstellar grain mantles (e.g., \citealt{Andersson+vanDishoeck+2008,Oberg+etal+2009a}).
Therefore, we have made improvements to incorporate the active bulk chemistry in the three-phase model, in which bulk diffusion is driven by the diffusion of water molecules in the ice (e.g., \citealt{Garrod+2013,Ruaud+etal+2016}). 
Additionally, we set the diffusion energy in ice mantles $E_{\rm diff}^{m}(i) = E_{\rm diff}^{m}$(H$_2$O) for all the species (except H, H$_2$, C, N, and O) with $E_{\rm diff}^{m}(i) < E_{\rm diff}^{m}$(H$_2$O), following the treatment of \citealt{Ruaud+etal+2016}.
The gas-phase network used in this work is based on the kinetic database for astrochemistry 2014 (KIDA 2014, \citealt{Wakelam+etal+2015}), and the grain network is the one presented by \cite{Ruaud+etal+2015,Ruaud+etal+2016}.
The grain network includes swapping reactions between species on surfaces and mantles and many reactions in mantles following \cite{Garrod+2013}.
Since the reaction-diffusion competition not only affects CO$_2$ formation on grains but also all species formed via reactions with a barrier, we use the competition mechanism for all two-body reactions in surfaces and mantles (see \citealt{Ruaud+etal+2016}).
The Eley-Rideal and van der Waals complex-induced reaction mechanisms play an important role in the formation of some COMs (\citealt{Ruaud+etal+2015}), so the grain surface network in this work also includes the new surface mechanisms (van der Waals complexes and Eley-Rideal mechanisms).

Besides some reaction mechanisms, we have also made some modifications to certain reactions in the chemical network.
The gas reaction rate coefficient of NH$_2$ + H$_2$CO $\rightarrow$ NH$_2$CHO + H was taken to be $2.6{\times}10^{-12}(T/300 {\rm K})^{-2.1}{\rm exp}(-26.9 {\rm K}/T)$ cm$^{3}$s$^{-1}$ following \cite{Barone+etal+2015}'s new insights.
To treat self- and mutual- shielding, we used the analytical approach of \cite{Draine+Bertoldi+1996} for H$_2$, tabulated rates for CO and N$_2$ photodissociation (e.g., \citealt{Visser+etal+2009,Li+etal+2013}).
Various studies on species on water ice have reported diffusion-desorption ratios from 0.3 to 0.6 (\citealt{Ruffle+Herbst+2000,Karssemeijer+Cuppen+2014,Minissale+etal+2016a,He+etal+2018}).
We assume $E_{\rm diff}^{s} = 0.4{\times}E_{\rm des}^s$ (except for H atoms; \citealt{Garrod+Herbst+2006,Ruaud+etal+2016,Aikawa+etal+2020}) and $E_{\rm diff}^{m} = 0.8{\times}E_{\rm des}^m$ (\citealt{Ruaud+etal+2016}) for surface and mantle species, respectively.
The diffusion barrier of H atom was adopted as 230 K, which is identical to its value in some previous work (\citealt{Al-Halabi+vanDishoeck+2007,Hama+etal+2012,He+etal+2018}).
We used a direct measurement of binding energy for atomic oxygen on dust grain surfaces of $\sim 1660$ K (\citealt{He+etal+2015}), which differs from some previous works (\citealt{Garrod+Pauly+2011}).  

\begin{table}
  \begin{center}
  \caption[]{Initial Abundances with Respect to Total Hydrogen in the Fiducial Model \label{InitialAbundances}}
  \setlength{\tabcolsep}{12pt}
  \small
  \begin{tabular}{llll}
    \hline
    \hline\noalign{\smallskip}
  Species & Abundance\tablefootmark{1} & Species & Abundance\tablefootmark{1}\\
    \hline\noalign{\smallskip}
  H & 3.33(-01) &Si$^+$ & 9.74(-09) \\
  H$_2$ & 3.33(-01) & Fe$^+$ & 2.74(-09) \\
  He & 9.00(-02) & Mg$^+$ & 1.09(-08) \\
  C$^+$ & 7.30(-05) & Cl$^+$ & 1.00(-09) \\
  N & 2.14(-05) & Na$^+$ & 2.25(-09) \\
  O & 1.76(-04) & P$^+$ & 2.16(-10) \\
  S$^+$ & 9.14(-08) & F & 6.68(-09) \\
    \hline
  \end{tabular}
  \tablefoot{\tablefoottext{1}{$A(B) = A{\times}10^B$}}
  \end{center}
\end{table}
We also modified some non-thermal desorption reactions, including cosmic-ray desorption, photodesorption, cosmic-ray-induced photodesorption, and chemical desorption reactions.
We used a new cosmic-ray desorption method following \cite{Kalvans+2018}, and the cosmic-ray flux was taken to be $10^4$ photons cm$^{-2}$ s$^{-1}$ (\citealt{Shen+etal+2004}). 
We added photodesorption and cosmic-ray-induced photodesorption processes following the approach of \cite{Kalvans+etal+2017}.
For CH$_4$, CO, CO$_2$, CH$_3$OH, N$_2$, NH$_3$, and H$_2$O molecules, we adopted experimental photodesorption yields (\citealt{Oberg+etal+2009b,Oberg+etal+2009c,Bertin+etal+2013,Martin-Domenech+etal+2015,Dupuy+etal+2017,Kalvans+2018}), while we set default photodesorption yields for other species of $10^{-4}$.
Reactive (chemical) desorption was also taken into account, simulating the ejection of products of exothermic surface reactions.
Experiments showed that the chemical desorption efficiency for partially hydrogenated reactions is very low on water ice surfaces (\citealt{Minissale+etal+2016b,Chuang+etal+2018}), which is consistent with theoretical calculations through Rice-Ramsperger-Kassel (RRK) theory (\citealt{Garrod+etal+2007}).
We adopted a generic value of 0.03 for chemical desorption efficiency for all exothermic surface reactions (\citealt{Garrod+etal+2007,Kalvans+2021}).
\cite{Andersson+vanDishoeck+2008} estimated that the probability of UV photons being absorbed by one ice monolayer is approximately 0.007, indicating that photodissociation reactions in the ice mantle should be taken into account.
\cite{Kalvans+2018} determined an average the solid/gas photodissociation coefficient ratio of $\sim0.3$, so we applied this value to icy mantles.
For the dust grain, we assume a canonical radius of $a = 0.1~\mu$m, a material density of $\rho_d = 3$ g cm$^{-3}$, an average albedo of 0.6, and a dust-to-gas mass ratio of 0.01 (e.g., \citealt{Aikawa+etal+2020,Kalvans+2021}).
The chemical parameters are listed in Table~\ref{BasicParameters}.
For the initial elemental abundances in the fiducial model, we assumed the so-called low-metal values, which are identical to those in \cite{Kalvans+2021}, except for hydrogen (\citealt{Garrod+etal+2008}).
We assumed gas-phase species were in atomic or ionized form, except for hydrogen, which was in molecular form.
The initial abundances in the fiducial model are given in Table~\ref{InitialAbundances}.

\section{Results}
\label{sect_Results}
\subsection{Physical Evolution in the Fiducial Model}
\begin{figure*} 
  \centering
  \includegraphics[width=\hsize, angle=0]{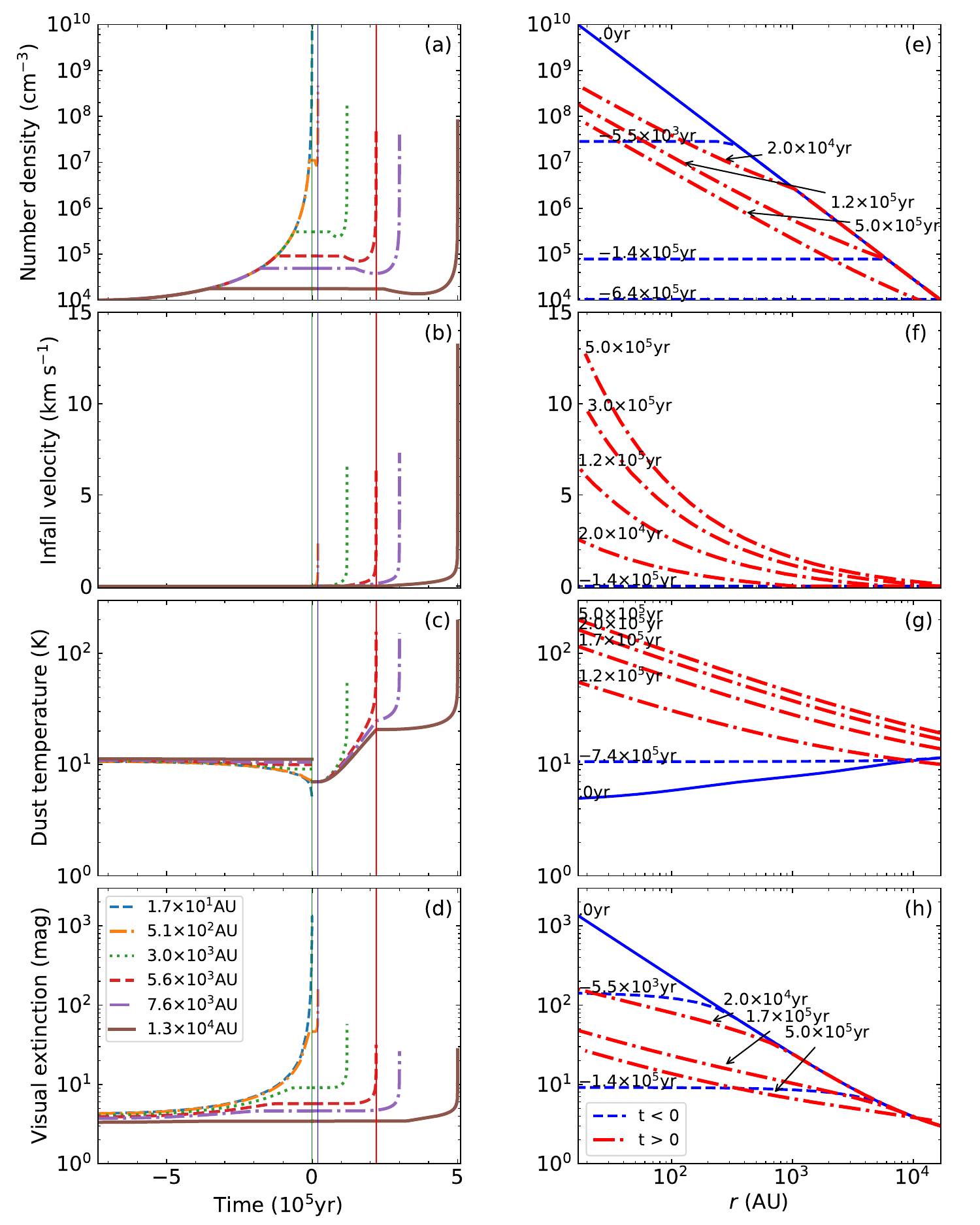}
  \caption{Physical parameters (i.e, number density, infall velocity, dust temperature, and visual extinction) varying with time and radius in the fiducial model. (a-d) Temporal variation of these physical parameters in specific gas parcels which initially locate at $r = 16.9$ AU, $5.1{\times}10^2$ AU, $3.0{\times}10^3$ AU, $5.6{\times}10^3$ AU, $7.6{\times}10^3$ AU, and $1.3{\times}10^4$ AU. 
  The three vertical lines (i.e., green, blue, and red line) indicate the initial moment of infall motion (i.e., $t = 0$), the beginning moment of warm-up (i.e., the birth of the protostar, $t = 2{\times}10^4$ yr), and the end of warm-up phase (i.e., $t = 2.2{\times}10^5$ yr), respectively. (e-h) Radial distributions of these physical parameters at some specific times. The times shown are relative to the initial moment of infall motion.} 
  \label{PhysicalParameters}
\end{figure*}

Several physical parameters dominate chemical processes in cloud cores, including number density, gas temperature, dust temperature, visual extinction, and cosmic-ray ionization rate.
The number density, infall velocity, dust temperature, and visual extinction in the fiducial model vary with radius and time are shown in Fig.~\ref{PhysicalParameters}.
Here, we define the moment of the prestellar core infall as $t = 0$.
The left panels show these physical parameters as a function of time at specific gas parcels, while the right panels show the radial distributions of these physical parameters at some specific times.

During the prestellar stages (i.e., $t < 0$), the number density at each radius increases with time before reaching a value in a singular isothermal sphere (\citealt{Rawlings+etal+1992}). 
Subsequently, the number density at this position stays constant until the gas parcel begins to infall (i.e., $t = 0$).
The number density in the outermost layer keeps at 10$^4$ cm$^{-3}$, while it in the innermost layer increases from 10$^4$ to 10$^{10}$ cm$^{-3}$.
Although the density at each position gradually increases, all gas parcels remain stationary.
The dust temperature at different radii decreases over time due to fewer UV photons heating the dust grain with increasing visual extinction.
Generally, the dust temperature keeps at approximately 10 K at any radius, bing higher outwards.
The visual extinction at each radius gradually increases to the corresponding maximum value in the prestellar phase, until the gas parcel begins to infall.
The visual extinction in the innermost layer increases from 4.3 to 1348 mag, while the visual extinction in the outermost layer is stable at 3.0 mag. 

When the inside-out collapse begins, the material in the regions swept by the expansion wave starts to infall toward the center. 
Additionally, the closer to the center, the faster the material falls. 
In the Lagrangian coordinates (see left panels in Fig.~\ref{PhysicalParameters}), the number densities of specific gas parcels keep constants before infall occurs, then gradually increase until gas parcels fall into the central protostar (i.e., $r < r_{\rm in}$ = 16.9 AU).
However, in the Eulerian coordinates (see right panels in Fig.~\ref{PhysicalParameters}), the number density at a fixed radius remains constant, then gradually decreases after the expansion wave passes by.
The material at a fixed radius falls increasingly faster over time.
During the isothermal collapse phase (i.e., $t < 2{\times}10^4$ yr), gas and dust temperatures are fixed at 7 K.
When the protostellar envelope enters the warm-up phase, the temperature in the innermost layer gradually increases from the initial 7 K to 200 K and remains there.
The visual extinction exhibits characteristics that vary with time and radius, which is similar to those of number densities in both Lagrangian and Eulerian coordinates.
\subsection{Chemical Evolution in the Prestellar Core}
\begin{figure} 
  \centering
  \includegraphics[width=\hsize, angle=0]{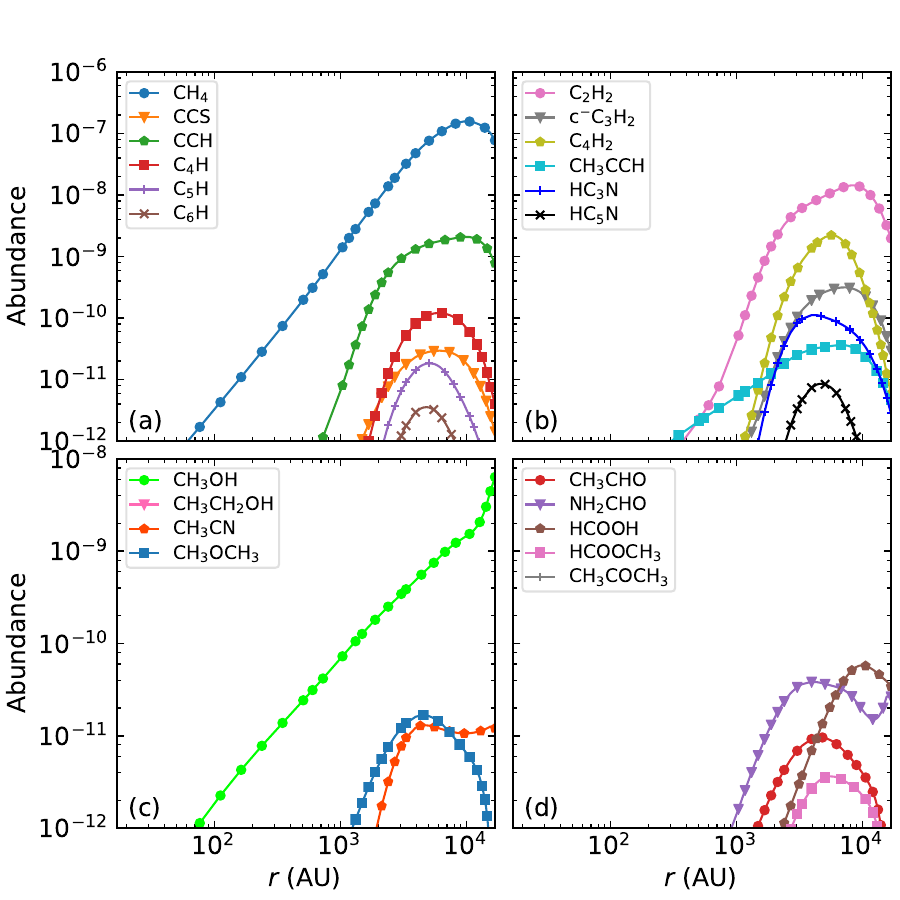}
  \caption{Radial distributions of some gaseous carbon-chain molecules (a-b) and COMs (c-d) at the end of the modified free-fall collapse (i.e., $t = 0$) in the fiducial model.
  } 
  \label{CCMsAndCOMsInPrestellarCores}
\end{figure}
Since chemical compositions in the protostellar envelope inherits from the prestellar core, we have to first investigate the molecular distribution during the prestellar phase.  
Panel (a-b) in Fig.~\ref{CCMsAndCOMsInPrestellarCores} show the radial distribution of some canonical carbon-chain molecules detected in prototypical WCCC sources, including C$_n$H, C$_n$H$_2$, HC$_{2n+1}$N families ($n = 1, 2, \cdots$), and some precursors related to WCCC at the end of the prestellar phase.
Most carbon-chain molecules are concentrated around the position of $4{\times}10^3$ AU, implying the emission of these molecules mainly comes from the outer regions of the prestellar core.
Carbon-chain molecules are rapidly adsorbed on the dust grain (adsorption timescale of $\sim$ 1 yr) due to high densities near the prestellar center, while their abundances are low near the peripheries due to strong UV photodissociation processes.  
The peak abundances of most of carbon-chain molecules, with the exception of CH$_4$, CCH, C$_2$H$_2$, and C$_4$H$_2$, are lower than $10^{-9}$.
For instance, the peak abundances of HC$_3$N and HC$_5$N molecules are close to the observed values in high-mass star-forming starless cores ($X$(HC$_3$N) $\sim 10^{-11}-10^{-10}$, and $X$(HC$_5$N) $\sim 10^{-12}-10^{-11}$; \citealt{Taniguchi+etal+2018}).
This suggests these abundant molecules observed in the protostellar core are not primarily inherited from the prestellar core, but instead another mechanism is responsible for their formation, known as the WCCC mechanism.

Panel (c-d) in Fig.~\ref{AbundancesOfCOMsInProtostellarCores} show the radial distributions of some saturated COMs detected in protostellar cores at the end of prestellar phase (i.e., $t = 0$).
These COMs are mainly produced on dust grain surfaces by the thermal hopping or quantum diffusion of radicals in the prestellar core.
For instance, the depleted CO molecule can be hydrogenated by hopping H atom, forming H$_2$CO and CH$_3$OH efficiently, which has been successfully confirmed by laboratory experiments (e.g., \citealt{Watanabe+etal+2003}).
Similar to the distributions of carbon-chain molecules, COMs are mainly concentrated at $r \sim 10^3-10^4$ AU.  
In high visual extinction regions ($A_{\rm V} > 5$ mag), gaseous COMs mainly decrease through adsorption processes, while those in outer layers are mainly destroyed by abundant UV photons.
Except for CH$_3$OH molecules, the peak abundances of most COMs are lower than $10^{-9}$, so COMs detected in protostellar envelopes are mainly not remnants from the prestellar core. 

\subsection{Chemical Evolution in the Protostellar Core}
\label{CheEvoInProtostellarCore}
\begin{figure*} 
  \centering
  \includegraphics[width=\hsize, angle=0]{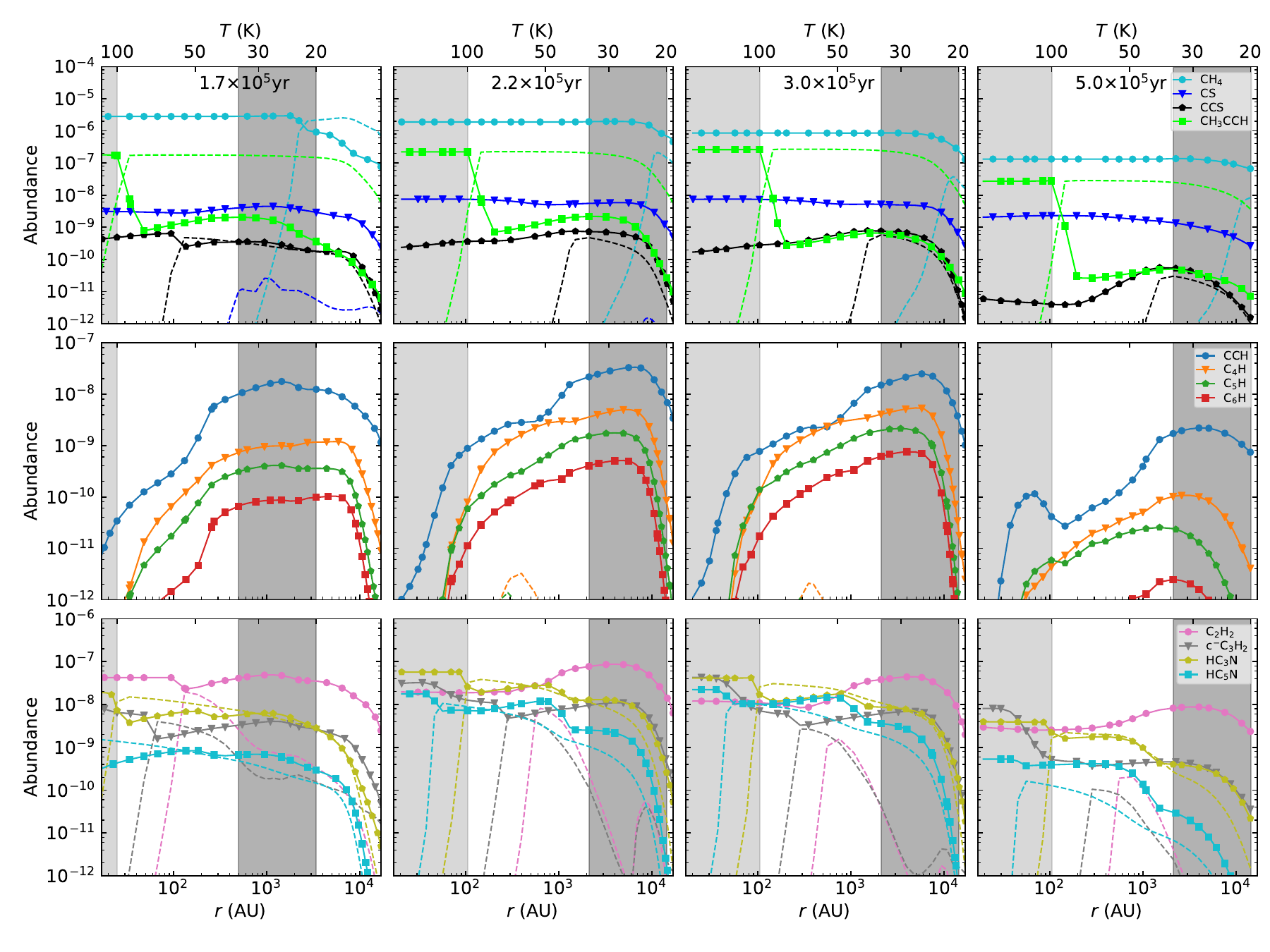}
  \caption{Radial distribution of simple (CH$_4$ and CS) and carbon-chain molecules at $t = 1.7{\times}10^5$, $2.2{\times}10^5$, $3.0{\times}10^5$, and $5.0{\times}10^5$ yr in the fiducial model. The light gray shadow covers the hot corino regions within the dust temperature $> 100$ K. The dark gray shaded areas represent the WCCC regions within the dust temperature from 20 to 35 K. The different molecules are distinguished by line colors and marks. The solid lines show the gaseous abundances, while the dashed lines depict the icy abundances.} 
  \label{AbundancesOfCCMsInProtostellarCores}
\end{figure*}
To understand whether the chemical hybrid sources are common phenomena, we investigate carbon-chain molecules and COMs in the fiducial model during the protostellar phase.
Figure~\ref{AbundancesOfCCMsInProtostellarCores} shows the radial distribution of unsaturated carbon-chain molecules and some molecules related to WCCC processes at some specific times.
Here, four specific times (i.e., $1.7{\times}10^5$, $2.2{\times}10^5$, $3.0{\times}10^5$, and $5.0{\times}10^5$ yr) represent different evolutionary stages.
Specifically, the first two moments represent the warm-up phase, and the last two moments represent the hot corino stage when the dust temperature at each radius remains constant.
The light gray shaded region in Fig.~\ref{AbundancesOfCCMsInProtostellarCores} is the hot corino region with dust temperature above 100 K, while the dark gray shaded region represents the WCCC region with dust temperature between 20-35 K (\citealt{Aikawa+etal+2020,Kalvans+2021}).

Firstly, we take the moment of $2.2{\times}10^5$ yr as an example to show the common distribution features of carbon-chain molecules at four moments.
There is a clear abundance jump for CH$_4$ molecule showed in the region within the dust temperature of $\sim 25$ K, corresponding to its sublimation temperature ($T_{\rm sub}$).
When the dust temperature is above 25 K, the solid CH$_4$ can quickly evaporate into the gas phase through thermal desorption.
Subsequently, gaseous CH$_4$ molecules react with C$^+$ to produce C$_2$H$_3^+$ or C$_2$H$_2^+$, which then react with e$^-$ to produce the shortest carbon-chain molecules, C$_2$H$_2$ and C$_2$H.
Long carbon-chain molecules are gradually produced through further reactions with C$^+$, H, and e$^-$ (for a detailed description, see \citealt{Hassel+etal+2008,Sakai+Yamamoto+2013}).
Therefore, carbon-chain molecules in the WCCC regions (i.e., $2.1{\times}10^3-1.4{\times}10^4$ AU) are more abundant relative to cold regions.
Abundance jump also appears for a few molecules, such as CH$_3$CCH ($T_{\rm sub} \sim 80$ K), C$_{n}$H$_2$ group, and HC$_{2n+1}$N group.
The abundances of carbon-chain molecules in most regions are higher than $10^{-10}$, indicating the common occurrence of carbon-chain molecules in protostellar cores.

Based on the radial distribution of carbon-chain molecules, they can be mainly divided into two types as shown in Fig.~\ref{AbundancesOfCCMsInProtostellarCores}.
The C$_n$H family represents the first type, which is characterized by an extended distribution and a small central dip around the protostar.
Numerous observations of CCH and C$_4$H have provided evidence for the widespread occurrence of this spatial distribution pattern (\citealt{Sakai+etal+2010,Imai+etal+2016,Oya+etal+2017,Murillo+etal+2018}). 
In hot corino regions (i.e., $r < 100$ AU), the destruction of C$_n$H molecules is dominated by the following reaction (the activation barrier is $\sim 950$ K): 
\begin{equation}
  \label{ReactionOfCnHWithH2}
  {\rm C}_n{\rm H} + {\rm H}_2 \rightarrow {\rm C}_n{\rm H}_2 + {\rm H},
\end{equation}
which accounts for a dip structure in the intensity profile toward the protostar.
The second type (including HC$_{2n+1}$N and C$_n$H$_2$ families) is characterized by central condensation, which is consistent with some observational evidences.
For instance, c-C$_3$H$_2$ molecules are distributed in more shielded inner envelopes (\citealt{Lindberg+etal+2017}), while HC$_5$N exhibits very high excitation energy levels (\citealt{Sakai+etal+2009,Taniguchi+etal+2023}).
In the lukewarm regions, HC$_{2n+1}$N molecules are mainly produced by the gaseous reactions (\citealt{Taniguchi+etal+2019})
\begin{equation}
  \label{ReactionOfC2HwithHCNType}
  {\rm C}_2{\rm H} + {\rm HC}_{2n-1}{\rm N} \rightarrow {\rm HC}_{2n+1} {\rm N + H}
\end{equation}
and 
\begin{equation}
  \label{ReactionOfC2nH2WithCN}
  {\rm C}_{2n}{\rm H}_2 + {\rm CN} \rightarrow {\rm HC}_{2n+1}{\rm N + H},
\end{equation}
while the precursors C$_{2n}$H$_2$ are produced by the reactions
\begin{equation}
  \label{ReactionOfC2nH2WithC2H}
  {\rm C}_{2n-2}{\rm H}_2 + {\rm C}_2{\rm H} \rightarrow {\rm C}_{2n}{\rm H}_2 + {\rm H}.
\end{equation}
These gaseous carbon-chain molecules are accreted onto the bulk and surface of ice mantles, which occurs at $25 < T_{\rm dust} < 100$ K.  
When the dust temperature approaches their sublimation temperature, they would be evaporated into the gas phase.
In the hot corino regions, due to the infall timescales of gas parcels of $\sim90$ yr are much less than the destruction timescales of the carbon-chain molecules, so their radial distributions show a flat structure in inner regions.
Due to high activation barriers between C$_n$H$_2$ and H$_2$ molecules, the central dip structure as C$_n$H family is absent in hot corino regions.

Comparing the distributions of these carbon-chain molecules at different moments, we find some clear differences in molecular abundances.
At the warm-up phase, the positions of the peak abundances for carbon-chain molecules gradually move outward, corresponding to WCCC regions.
At the hot corino phase, the peak abundances of carbon-chain molecules keep at the WCCC regions (i.e., $2{\times}10^3-1.4{\times}10^4$ AU).
As the protostellar evolution processes, the carbon-chain molecules at any radius gradually decrease, which is mainly because the CH$_4$ abundance near the cloud periphery is less abundant due to its rapid reaction with H$_3^+$.
At the late stages (e.g., $t = 5{\times}10^5$ yr), the spatial distributions of carbon-chain molecules in the protostellar envelope are strongly affected by the remnants from the prestellar core.
In general, WCCC is inclined to appear during the warm-up phase, which is consistent with some models constructed by \cite{Hassel+etal+2008} and \cite{Wang+etal+2019}.

\begin{figure*} 
  \centering
  \includegraphics[width=\hsize, angle=0]{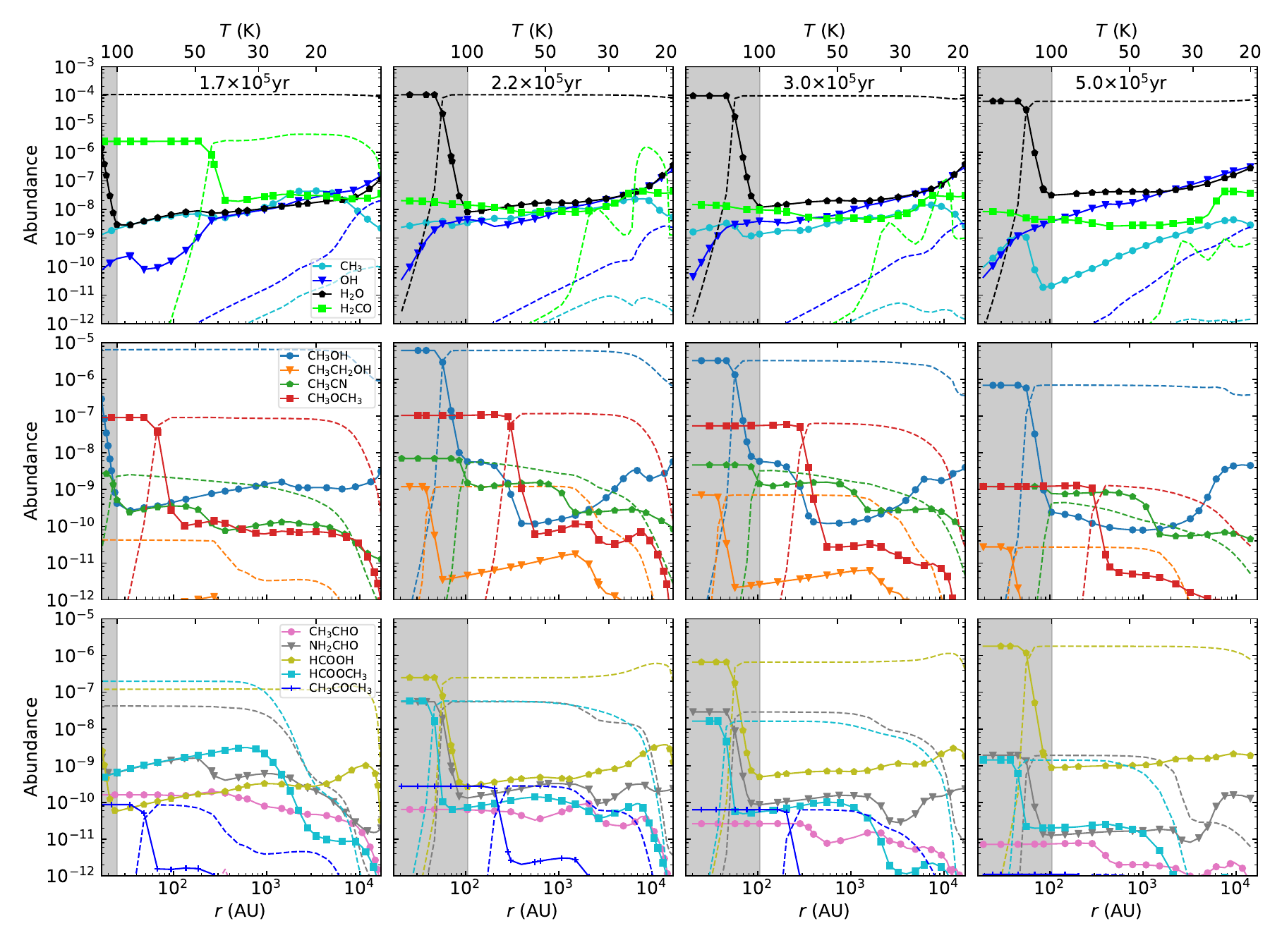}
  \caption{Radial distribution of simple molecules (CH$_3$, OH, H$_2$O, and H$_2$CO) and COMs at $t = 1.7{\times}10^5$, $2.2{\times}10^5$, $3.0{\times}10^5$, and $5.0{\times}10^5$ yr in the protostellar core, as in Fig.~\ref{AbundancesOfCCMsInProtostellarCores}.} 
  \label{AbundancesOfCOMsInProtostellarCores}
\end{figure*}
Figure~\ref{AbundancesOfCOMsInProtostellarCores} shows the spatial distribution of saturated COMs and some molecules related to hot corino chemistry at some specific times, identical with in Fig.~\ref{AbundancesOfCCMsInProtostellarCores}.
The distribution characteristics of COMs are significantly diverse from those of carbon-chain molecules in Fig.~\ref{AbundancesOfCCMsInProtostellarCores}, due to different formation pathways.
Since COMs are mainly produced by surface reactions on dust grains, the dust temperature is the most important factor to determine their radial distributions.

Firstly, we take the time of $2.2{\times}10^5$ yr as an example to depict the similar distribution characteristics of COMs at these four moments.
The abundances of COMs in hot corino regions are extremely higher than $10^{-10}$, indicating the common occurrence of COMs in protostellar cores.
Hot corino chemistry starts with the successive hydrogenation of CO molecules on the surfaces of dust grains to form CH$_3$OH, which serves as a fundamental COM and a precursor for more complex COMs (\citealt{Oberg+Bergin+2021}). 
In hot corino regions (i.e., $r < 100$ AU; $T_{\rm dust} > 100$ K), most saturated COMs in ice mantles are evaporated into the gas phase by the thermal desorption. 
Infall timescales of gas parcels through hot corino regions ($\sim 90$ yr) are less than destruction timescales of COMs (e.g., $t_{\rm des} \sim5{\times}10^5$ yr for CH$_3$OH), so most COMs exhibit a flat profile.
Additionally, saturated COMs are mainly distributed in the regions within 100 AU near the central protostar, which is consistent with a few observations toward protostars.
For example, some observations toward some hybrid sources, such as B335, L483 (\citealt{Imai+etal+2016,Oya+etal+2017,Bouvier+etal+2020}) showed narrow spatial distributions for saturated COMs.
An abundance jump has been observed for CH$_3$OH and H$_2$CO in several low-mass protostellar sources (e.g., \citealt{Maret+etal+2005,Kristensen+etal+2010}), as shown in the fiducial model.

There are some clear variations in molecular abundances at these four times.
During the warm-up phase, the peak abundances of COMs gradually increase with time due to the expanding hot corino regions, where most COMs in ice mantles are evaporated into gas-phase.
When the infall timescales are larger than the evaporation timescales of COMs, the molecular abundance can form a rapidly rising profile at $t = 1.7{\times}10^5$ yr, without generating a flat feature in inner regions as at other moments in Fig.~\ref{AbundancesOfCOMsInProtostellarCores}.
At the late stages, the material near the outer boundary of protostellar envelope begins to infall towards the protostar.
The COMs and their precursors at the periphery are rapidly destroyed by UV photons and secondary photons (destruction timescale is $1{\times}10^4$ yr for CH$_3$OH), leading to low abundances of COMs (i.e., $\le 10^{-9}$) in gas phase and ice mantles.

The chemical survey toward young stellar objects in the Perseus molecular cloud complex indicates that the hybrid sources could be a widespread occurrence.
Additionally, infrared observations by \cite{Graninger+etal+2016} also suggest an indispensible evidence for the hybrid sources, by observing simultaneously presence of CH$_4$ and CH$_3$OH ices. 
The emission of carbon-chain molecules is more extended compared to that of COMs in the protostellar core, as demonstrated in hybrid sources (\citealt{Oya+etal+2017,Taniguchi+etal+2021a}).
The fiducial model predicts abundant saturated COMs near the innermost protostellar envelopes at $2.2{\times}10^5$ yr, corresponding to hot corino chemistry.
Simultaneously, the model also predicts abundant unsaturated carbon-chain molecules in WCCC regions due to the sublimation of methane.
Therefore, unsaturated carbon-chain molecules and saturated COMs can coexist, which has been observed in some protostellar cores (e.g., B335, \citealt{Imai+etal+2016,Imai+etal+2019}; L483, \citealt{Oya+etal+2017,Jacobsen+etal+2019,Agundez+etal+2019}).
Since the emissions of carbon-chain molecules and COMs are spatially separated in Figs.~\ref{AbundancesOfCCMsInProtostellarCores} and \ref{AbundancesOfCOMsInProtostellarCores}, there is no clear statistical correlation between their column densities in large sample observations (e.g., \citealt{Lindberg+etal+2016}).

\subsection{Comparison with Observations}
\label{sect_ComparisonWithObservations}
Generally, the CH$_4$/H$_2$O abundance in bulk ice around low-mass young stellar objects (YSOs) are nearly constant within a narrow range (2-8\%), basing on the `Cores to Disks' Spitzer Spectroscopic Survey conducted by \cite{Oberg+etal+2008}.
At early warm-up phase (e.g., $t \sim 1.7{\times}10^5$ yr), the peak abundance of solid CH$_4$ and solid H$_2$O is close to $3{\times}10^{-6}$ and $1{\times}10^{-4}$, respectively, so their ratio equals to 3\% aligning with observed values.
At hot corino stages (e.g., $3.0{\times}10^5$ yr, see in Fig.~\ref{AbundancesOfCCMsInProtostellarCores}), most CH$_4$ ice evaporates into the gas phase, which may explain why approximately half of low-mass YSOs lack a solid CH$_4$ signal (\citealt{Oberg+etal+2008}).

The abundances of carbon-chain molecules and COMs in the fiducial model differ from the results in some works such as \cite{Aikawa+etal+2008,Aikawa+etal+2020}.
The variation between our three-phase model and the two-phase or multilayered ice mantle model used in their works is one source for the differences.
However, other factors may amplify chemical differences as well, such as the physical model (inside-out collapse model vs radiation hydrodynamic model).
For instance, the peak abundance of NH$_2$CHO in the fiducial model is significantly higher, by more than one order of magnitude, compared with the simulated value in \cite{Aikawa+etal+2020}.
On the other hand, we did not consider several dynamical changes, such as the molecular outflow and disk formation, so our models may contradict some observed results dominated by above processes.
For instance, \cite{Sakai+etal+2006} detected COMs even at a very early stage in protostellar evolution toward the low-mass protostar NGC 1333 IRAS 4B.

\begin{table}
  \begin{center}
  \caption[]{Observed and Simulated Abundances in L483 \label{HybridL483}}
  \setlength{\tabcolsep}{16pt}
  \small
  \begin{tabular}{lcc}
    \hline
    \hline
  Species & Observations\tablefootmark{1} & Simulations\tablefootmark{1,2} \\
    \hline
  CN & 5.51(-09)\tablefootmark{a} & 4.13(-08) \\
  CS & 2.89(-09)\tablefootmark{a} & 8.20(-09) \\
  SO & 5.00(-09)\tablefootmark{a} & 6.44(-09) \\
  SiO & 6.50(-12)\tablefootmark{a} & \textbf{3.71(-10)} \\
  HCN & 2.97(-09)\tablefootmark{a} & \textbf{7.29(-08)} \\
  HNC & 5.95(-09)\tablefootmark{a} & 3.13(-08) \\
  HCO$^+$ & 7.53(-09)\tablefootmark{a} & 1.29(-08) \\
  CCH & 1.28(-08)\tablefootmark{a} & 3.95(-08) \\
  CCS & 1.22(-10)\tablefootmark{a} & 6.53(-10) \\
  HNCO & 1.77(-09)\tablefootmark{b} & \textbf{6.05(-08)} \\
  c-C$_3$H$_2$ & 5.25(-09)\tablefootmark{a} & 1.11(-08)\\
  CH$_3$OH & 7.31(-09)\tablefootmark{a} & 9.66(-09) \\
  CH$_3$CHO & 3.60(-10)\tablefootmark{b} & 2.81(-10) \\
  HCOOCH$_3$ & 1.07(-08)\tablefootmark{b} & 8.03(-09) \\
  SO$_2$ & 5.88(-03)\tablefootmark{c} & \textbf{4.43(-04)} \\
  H$_2$CS & 1.18(-03)\tablefootmark{c} & 8.20(-03) \\
  HC$_3$N & 2.94(-02)\tablefootmark{c} & 1.45(-02) \\
  NH$_2$CHO & 5.88(-04)\tablefootmark{c} & \textbf{1.30(-02)} \\
  CH$_3$OCH$_3$ & 4.71(-03)\tablefootmark{c} & 3.79(-02) \\
  CH$_3$CH$_2$OH & 5.88(-03)\tablefootmark{c} & \textbf{4.70(-05)} \\
    \hline\noalign{\smallskip}
  \end{tabular}
  \tablefoot{
   \tablefoottext{1}{$A(B)=A{\times}10^B$}.
   \tablefoottext{a}{The fractional abundance relative to H$_2$ is from \cite{Agundez+etal+2019}, and the spatial resolution is $\sim 5000$ AU.}
   \tablefoottext{b}{The fractional abundance relative to H$_2$ is from \cite{Oya+etal+2017}, and the spatial resolution is $\sim 100$ AU.}
   \tablefoottext{c}{The fractional abundance relative to CH$_3$OH is from \cite{Jacobsen+etal+2019}, and the spatial resolution is $\sim 20$ AU.}
   \tablefoottext{2}{The simulated beam-average fractional abundances have already been smoothed to their respective spatial resolutions. Boldface indicates a disagreement of more than one order of magnitude between the observed and modeled value.}
   }
  \end{center}
\end{table}
Among the hybrid sources identified through interferometer observations, the low-mass protostellar source L483 stands out as a representative example.
It is located at a distance of 200 pc, as reported in studies such as \cite{Dame+Thaddeus+1985} and \cite{Jorgensen+etal+2002}.
Previously, L483 was categorized as a WCCC source due to its bright emission from carbon-chain molecules such as HC$_5$N (e.g., \citealt{Hirota+etal+2009}). 
However, high-resolution observations recently revealed an active hot corino chemistry near the protostar (e.g., \citealt{Oya+etal+2017,Jacobsen+etal+2019}).
Meanwhile, it exhibits an exceedingly rich chemical composition, including several new identified interstellar molecules such as HCCO (\citealt{Agundez+etal+2015a}), NCCNH$^+$ (\citealt{Agundez+etal+2015b}), NCO (\citealt{Marcelino+etal+2018b}), etc.
Based on above considerations, we selected Class 0 object L483 as a reference for comparison with the fiducial model.
These molecules used for comparison are listed in Table~\ref{HybridL483}, with references from \cite{Oya+etal+2017,Agundez+etal+2019,Jacobsen+etal+2019}. 
The angular resolutions in these articles are 0.5$^{\prime\prime}$, 25$^{\prime\prime}$, and 0.1$^{\prime\prime}$, corresponding to spatial resolutions of 100, 5000, and 20 AU, respectively.
For molecules involved in multiply works, we adopted their observed values according to the fitting goodness. 
In \cite{Agundez+etal+2019}'s work, we selected molecules with line widths higher than 1.0 km/s, which is mainly because molecules with low line widths primarily originate from its cold ambient cloud, rather than from warm or hot compact regions.

In the case of low spatial resolution, the abundances of centrally concentrated molecules can be severely diluted, so we need to calculate average column densities of these molecules.
To assess the column densities of molecules at each position along the line-of-sight, we used spherical protostellar envelopes.
Subsequently, by smoothing the column density at each position for different telescopes, we obtained the average column densities and fractional abundances for listed molecules in Table~\ref{HybridL483}. 
One method to quantity the overall goodness between simulations and observations is the ``mean confidence level'' (\citealt{Garrod+etal+2007,Hassel+etal+2008}).
The confidence level of specie $i$, designated by $\kappa_i$, indicates the level of agreement between simulated values $X_i$ and observed values $X_i^{\rm obs}$, thus:
\begin{equation}
  \kappa_i = 1 - erf \left(\frac{|{\rm log}(X_i)-{\rm log}(X^{\rm obs}_i)|}{\sqrt{2}\sigma}\right)
\end{equation}
where $erf$ is the error function, and $\sigma$ is the standard deviation.
We adopted quantitative standard of one order of magnitude between the observed and simulated values.
Therefore, the mean confidence level $\kappa$ is 0.317, when $\sigma = 1$.
We got the maximum $\kappa$ of 0.508, with the best-fitting time being $1.965{\times}10^5$ yr.

Another method to assess the global agreement is the fitting number, which represents the species that align with the observations within one order of magnitude.
At the fitting time, out of the 20 molecules analyzed, 14 exhibited successful fitting.
However, there were six molecules (SiO, SO$_2$, HCN, HNCO, NH$_2$CHO and CH$_3$CH$_2$OH) that did not fit well.
If we consider only unsaturated carbon-chain molecules and saturated COMs, 8 out of 10 molecules can be fitted well.
Overall, the fiducial model can reproduce canonical characteristics of WCCC and hot corino chemistry in hybrid source L483.

\section{Possible reasons for chemical differences between WCCC and hot corino sources}
\label{DependenceOnSomePhysicalParameters}
To figure out main factors leading to chemical differences between WCCC sources and hot corino sources, we investigate the dependence of carbon-chain molecules and COMs on several physical parameters.
We adopt different values of the visual extinction of ambient clouds ($A_{\rm V}^{\rm amb}$ = 0.1, 1, 2, 3), cosmic-ray ionization rate ($\zeta = 3{\times}10^{-18}$, $3{\times}10^{-17}$, $3{\times}10^{-16}$ s$^{-1}$; \citealt{Neufeld+etal+2010,Indriolo+etal+2010,Indriolo+McCall+2012,Neufeld+Wolfire+2017}), the maximum temperature ($T_{\rm max} =$ 150, 200, 250, 300 K), and the contraction timescale ($t_c = 7{\times}10^4$, $2{\times}10^5$, $5{\times}10^5$ yr; \citealt{Bernasconi+Maeder+1996}).
The values of the varied physical parameters are listed in Table~\ref{VariedParameters}.
The abundances of most unsaturated carbon-chain molecules and saturated COMs at late stage (i.e., $5{\times}10^5$ yr) are lower than $10^{-9}$, as shown in Figs.~\ref{AbundancesOfCCMsInProtostellarCores} and \ref{AbundancesOfCOMsInProtostellarCores}, showing weak WCCC and hot corino chemistry.
Therefore, we no longer consider the dependence of WCCC and hot corino chemistry on these physical parameters at $t = 5{\times}10^5$ yr.
\begin{table*}
  \begin{center}
  \caption{Varied Physical Parameters in All Models\label{VariedParameters}}
   \begin{tabular}{lc}
    \hline
    \hline\noalign{\smallskip}
  Visual extinction of ambient clouds ($A_{\rm V}^{\rm amb}$) & 0.1, 1, 2, 3 (mag)\\
  Cosmic-ray ionization rate ($\zeta$) & $3{\times}10^{-18}$, $3{\times}10^{-17}$, $3{\times}10^{-16}$ (s$^{-1})$\\
  Maximum temperature ($T_{\rm max}$) & 150, 200, 250, 300 (K)\\
  Contraction timescale ($t_{\rm cont}$) & $7{\times}10^4$, $2{\times}10^5$, $5{\times}10^5$ (yr)\\
  \hline\noalign{\smallskip}
  \end{tabular}
\end{center}
\end{table*}

\subsection{Dependence on Visual Extinction of Ambient Clouds}
\label{VisualExtinction}
\begin{figure*} 
  \centering
  \includegraphics[width=\hsize, angle=0]{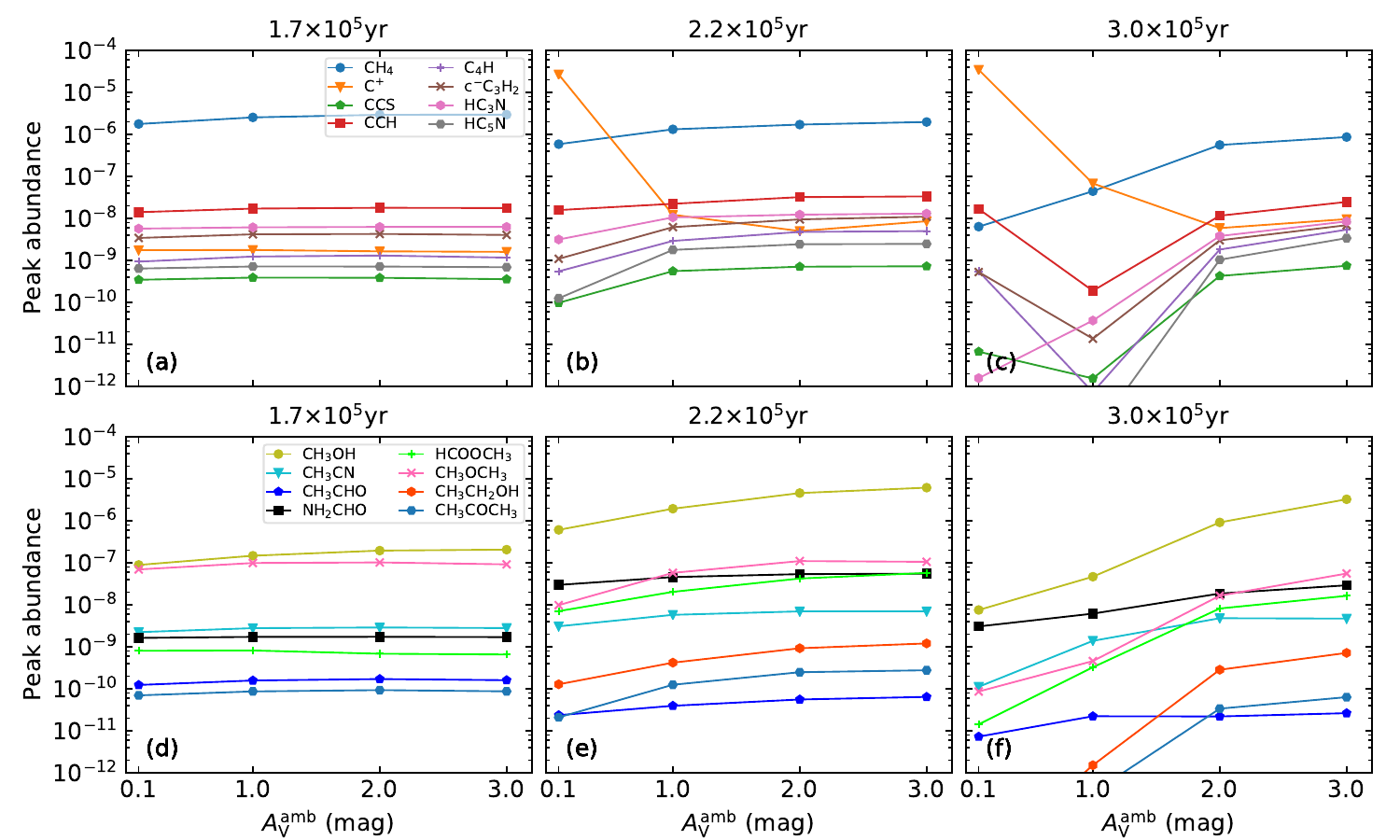}
  \caption{Peak abundances at specific regions vary with visual extinction of ambient clouds at some specific times. (a-c) Peak abundances of gaseous carbon-chain molecules around the CH$_4$ sublimation regions ($T$ = 20 - 35 K). (d-f) Peak abundances of gaseous COMs around the hot corino regions ($T \geq 100$ K).} 
  \label{VariationOfAv}
\end{figure*}
Numerous observations have indicated that interstellar radiation field and visual extinction of ambient clouds can drive the chemical differences among protostellar cores.
\cite{Spezzano+etal+2016} found that chemical variation can be influenced by different intensity of radiation from interstellar fields, as observed in the L1544 prestellar core.
Their findings revealed that photochemistry can maintain more C atoms in the gas phase, facilitating the carbon-chain production.
Another study from \cite{Sicilia-Aguilar+etal+2019} indicated IC 1396A is resembled with the hybrid sources in chemical features, in which WCCC was attributed to the UV irradiation. 
Some statistical observations suggested that WCCC sources are located at the cloud peripheries, while hot corino sources tend to be deeply embedded into dense filamentary clouds (e.g., \citealt{Higuchi+etal+2018,Lefloch+etal+2018}).
Recently, \cite{Bouvier+etal+2022} conducted statistical observations toward the OMC-2/3 filament, and found that the detection rate of hot corinos appears to be scarce compared to Perseus low-mass star-forming regions.
In general, the interstellar radiation field plays a significant role in chemical differences among the low-mass protostars.
Here we take the visual extinction of ambient clouds as an example, to explore the influence of the interstellar fields.
\begin{figure*} 
  \centering
  \includegraphics[width=\hsize, angle=0]{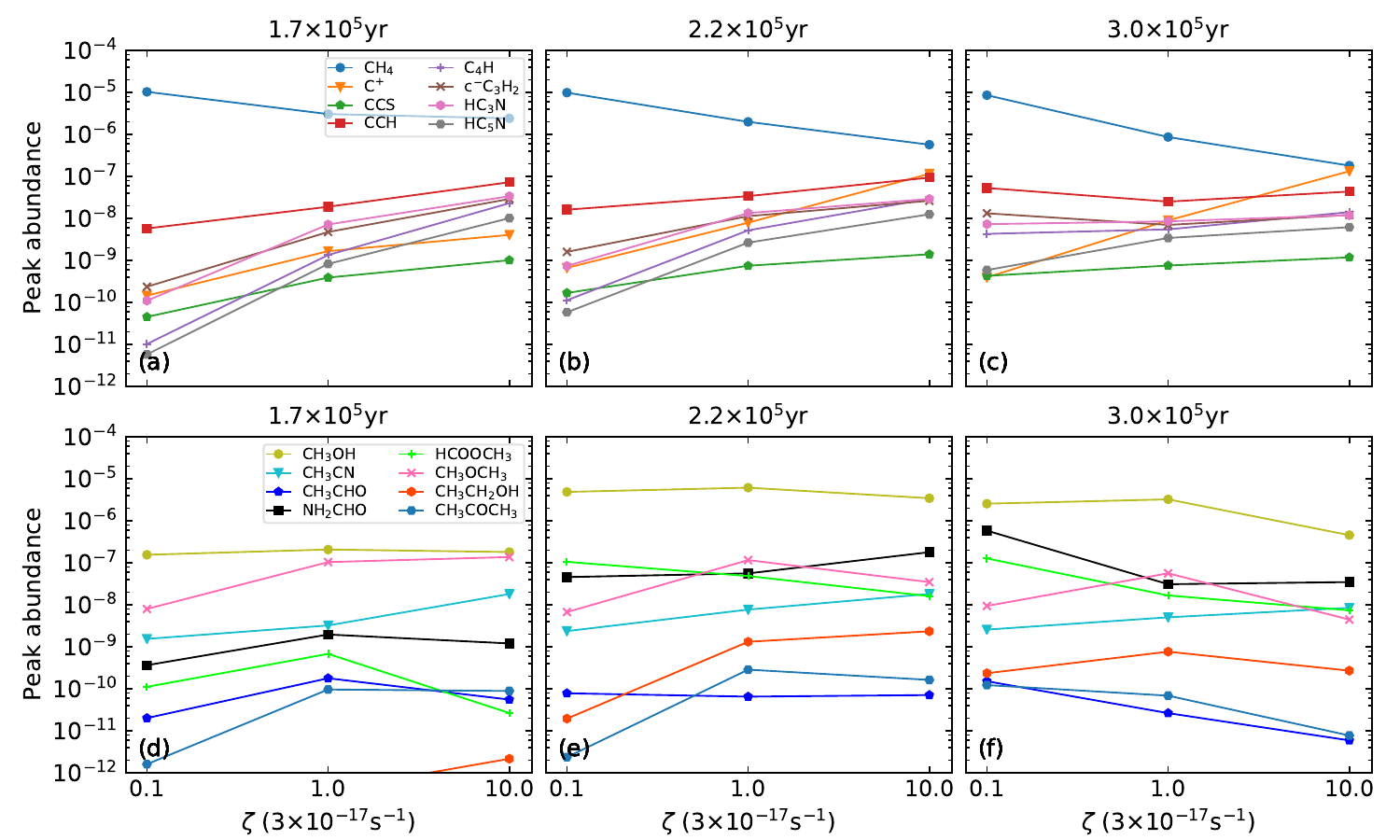}
  \caption{Peak abundances at specific regions vary with cosmic ray ionization rates at some specific times. (a-c) Peak abundances of gaseous carbon-chain molecules around the CH$_4$ sublimation regions. (d-f) Peak abundances of gaseous COMs around the hot corino regions.} 
  \label{VariationOfCRIR}
\end{figure*}

Figure~\ref{VariationOfAv} shows peak abundances of gaseous carbon-chain molecules around the CH$_4$ sublimation regions (i.e., WCCC regions within the dust temperature from 20 to 35 K) and gaseous COMs around the hot corino regions (i.e., the dust temperature of $\geq 100$ K) varying with the visual extinction of ambient clouds ($A_{V}^{\rm amb}$) at some specific moments.
Carbon-chain molecules primarily are formed in the gas phase, as shown in Fig.~\ref{AbundancesOfCCMsInProtostellarCores}, so the peak abundances of carbon-chain molecules at any time are not significantly affected by non-thermal desorption mechanisms (e.g., cosmic-ray desorption, photodesorption, and chemical desorption).
Since $A_{V}^{\rm amb}$ mainly affects chemical results by photodissociation reactions, which are not the major destruction pathways in inner regions, the peak abundances of carbon-chain molecules are insensitive to this factor at early warm-up stages (e.g., $t = 1.7{\times}10^5$ yr). 
At the time of $2.2{\times}10^5$ yr, the peak abundances of carbon-chain molecules slightly grow with increasing $A_{V}^{\rm amb}$, due to weak photodissociation processes.
At hot corino phase ($t > 2.2{\times}10^5$ yr), the peak abundances of carbon-chain molecules reach to the minimum value of $< 10^{-9}$ at $A_{\rm V}^{\rm amb} \sim 1.0$ mag.
At lower $A_{\rm V}^{\rm amb}$, strong UV photons quickly dissociate CO molecules, generating more C atoms and C$^+$ ions, which accelerates the formation of carbon-chain molecules through the WCCC mechanism mentioned in Sect.~\ref{CheEvoInProtostellarCore}.
This explains some observations that the c-C$_3$H$_2$ emission is effectively enhanced due to the UV radiation from a Herbig Be star (\citealt{Taniguchi+etal+2021b}), and that WCCC sources tend to be located at the cloud boundaries (e.g., \citealt{Higuchi+etal+2018,Lefloch+etal+2018}).
At $A_{\rm V}^{\rm amb} \geq 2.0$ mag, the production of C$^+$ ions primarily depends on secondary photons generated by cosmic rays due to the strong CO self-shielding, so the C$^+$ abundance remains at a low level. 
Meanwhile, the destruction of CH$_4$ and carbon-chain molecules is ineffective by photodissociation reactions, so their peak abundances slightly increase with increasing $A_{\rm V}^{\rm amb}$.

In contrast, COMs are mainly evaporated into the gas phase through the thermal desorption, as shown in Fig.~\ref{AbundancesOfCOMsInProtostellarCores}, so their peak abundances are not effectively affected by non-thermal desorption mechanisms.
At warm-up phase ($t = 2{\times}10^4-2.2{\times}10^5$ yr), the peak abundances of COMs slightly grow with increasing $A_{V}^{\rm amb}$, due to slow photodestruction rates.
At hot corino phase, COMs in hot corino regions are formed on the dust surfaces and subsequently evaporated into the gas phase through thermal desorption processes. 
At a high $A_{\rm V}^{\rm amb}$, less UV photons weaken the photodissociation of COMs and their precursors, producing high peak abundances, which is consistent with high detection rate of hot corino sources in Perseus low-mass star-forming regions (\citealt{Yang+etal+2021}) compared to in the OMC-2/3 filament reported by \cite{Bouvier+etal+2022}.
As $A_{\rm V}^{\rm amb}$ increases from 0.1 to 3.0 mag, the peak abundances of most COMs increase by about two orders of magnitude.
This explains statistical observations indicating that hot corino sources are predominantly located in inner regions (e.g., \citealt{Higuchi+etal+2018,Lefloch+etal+2018}).
Overall, the variation of $A_{\rm V}^{\rm amb}$ can lead to the chemical differences between WCCC sources and hot corino sources.

\subsection{Dependence on Cosmic Rays Irradiation}
\label{CosmicRay}
The cosmic ray ionization rate is another important physical parameter that can influence the chemical differences among low-mass protostars.
\cite{Kalvans+2021} found that the WCCC features in protostellar cores may originate from regions irradiated by cosmic rays, while the abundances of COMs have no clear correlation with radiation.
They utilized UDFA 12 network launched by \cite{McElroy+etal+2013}, which differs from the KIDA 2014 network.
The potential impact of differences between these two networks on our simulated results is discussed in Sect.~\ref{InfluenceOfGasPhaseNetwork}.
On the other hand, they employed a single-point physical model that cannot accurately calculate the peak abundances observed in the hot corino regions.
Therefore, we reinvestigate the impact of cosmic ray ionization rates on peak abundances of carbon-chain molecules and COMs.
\begin{figure*} 
  \centering
  \includegraphics[width=\hsize, angle=0]{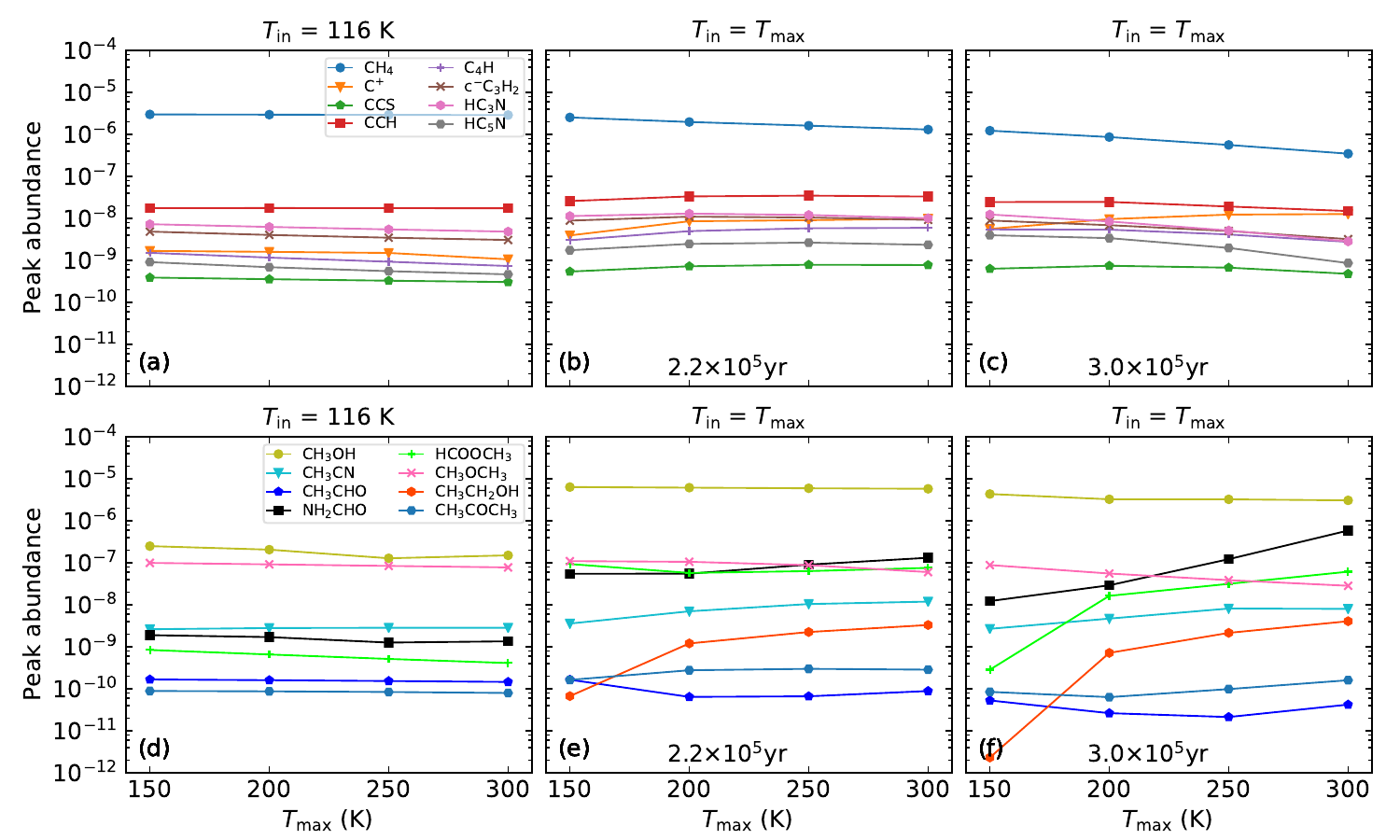}
  \caption{Peak abundances at specific regions vary with the maximum temperature at some fixed temperatures in the innermost layer or times. (a-c) Peak abundances of gaseous carbon-chain molecules around the CH$_4$ sublimation regions. (d-f) Peak abundances of gaseous COMs around the hot corino regions.} 
  \label{VariationOfTmax}
\end{figure*}

Figure~\ref{VariationOfCRIR} shows the peak abundances of gaseous carbon-chain molecules and COMs varying the cosmic ray ionization rates at some specific times.
Since the peak abundances of carbon-chain molecules and COMs are affected differently by cosmic ray ionization rates, we analyze their responses to this physical parameter separately.
The peak abundances of carbon-chain molecules show a significant increase with enhancing cosmic rays, except for the reduced precursor molecule - CH$_4$. 
As the cosmic ray ionization rate increases, the dissociation of CO and CH$_4$ molecules accelerates, producing more C atoms. 
Subsequently, C atoms are ionized into C$^+$ by the cosmic-ray-induced photoreactions and react with CH$_4$ molecules through the WCCC mechanism, leading to more abundant unsaturated carbon-chain molecules and lower peak abundances of CH$_4$.
At $\zeta=3{\times}10^{-16}$ s$^{-1}$, the peak abundances of most carbon-chain molecules are higher than $10^{-9}$, indicating that cosmic rays contribute to WCCC processes. 

The influence of cosmic ray ionization rate on the peak abundances of COMs is relatively complex.
This is mainly because COMs and their precursors are affected by various reaction types, such as cosmic ray desorption, cosmic-ray-induced photodesorption, cosmic-ray-induced photodissociation in gas phase and ice mantle, and cosmic-ray reactions.
Additionally, the formation pathways of COMs vary significantly, so we cannot give a conclusion regarding to the impact of cosmic rays on the peak abundances of COMs.
One can study how the peak abundance of a specific molecule vary with the cosmic ray ionization rate, by analyzing its main formation and destruction reactions in the chemical network. 
The variation of the cosmic ray ionization rate can lead to differences in WCCC, but the response of hot corino chemistry to its variation is highly dependent on the species.

\subsection{Independence on Maximum Temperatures in Hot Corinos}
\label{MaximumTemperatures}
We explored how some physical parameters at warm-up phase of the protostar as well, such as the maximum temperature and contraction timescale, influence WCCC and the hot corino chemistry.
Some observations indicated that the abundance and detection number of COMs statistically do not have a direct relation with protostellar properties such as $L_{\rm bol}$ and $T_{\rm bol}$ (e.g, \citealt{Higuchi+etal+2018,Yang+etal+2021}), implying that the protostellar properties may not lead to differences in COMs among protostars.
On the other hand, the luminosity of prototypical WCCC sources (e.g., 1.9 $L_{\odot}$ in L1527; 1.8 $L_{\odot}$ for IRAS 15398) tends to lower than that of prototypical hot corino sources, for instance, 9.1 $L_{\odot}$ in NGC 1333 IRAS4A and 22 $L_{\odot}$ in IRAS 16293 (\citealt{Froebrich+2005,Crimier+etal+2010,Kristensen+etal+2012,Jorgensen+etal+2013,Karska+etal+2013}).
This suggests that the chemical differentiations between hot corino sources and WCCC sources may be influenced by the protostellar luminosity, which affects some physical parameters at warm-up phase such as maximum temperature.
\begin{figure*} 
  \centering
  \includegraphics[width=\hsize, angle=0]{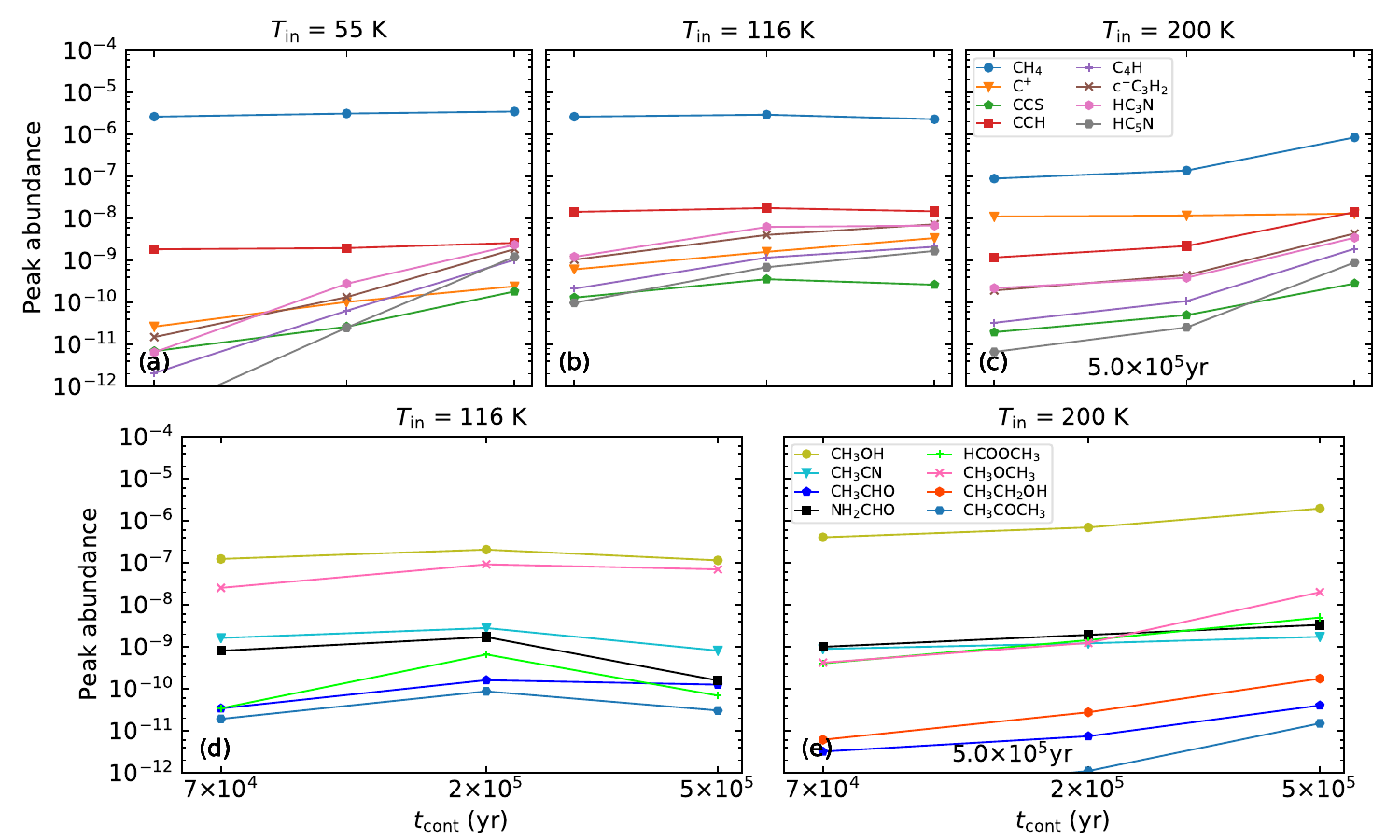}
  \caption{Peak abundances at specific regions vary with the contraction timescale at some fixed temperatures in the innermost layer. (a-c) Peak abundances of gaseous carbon-chain molecules around the CH$_4$ sublimation regions. (d-e) Peak abundances of gaseous COMs around the hot corino regions.} 
  \label{VariationOfContractionTime}
\end{figure*}

Considering above observed results, we attempt to investigate the influence of the maximum temperature at warm-up phase ($T_{\rm max}$) on the peak abundances of carbon-chain molecules and COMs, as shown in Fig.~\ref{VariationOfTmax}.
Note that the peak abundances of carbon-chain molecules and COMs are compared at the same temperature in the innermost layer ($T_{\rm in}$) at the first moment, while they are compared at the same times at hot corino stages (i.e., $t \geq 2.2{\times}10^5$ yr), as shown in Fig.~\ref{VariationOfAv}.
At warm-up moments, the time to reach the same $T_{\rm in}$ differs for different $T_{\rm max}$, so one expects variations in the simulated results.
At warm-up and hot corino stages, the peak abundances of carbon-chain molecules are insensitive to $T_{\rm max}$.
Even at a high temperature, HC$_{2n+1}$N families remain abundant, which are consistent with radio observations toward some high-mass YSOs conducted by \cite{Taniguchi+etal+2019}.

COMs can be categorized into two groups based on the variations in their peak abundances with $T_{\rm max}$.
The peak abundances in the first group (e.g., NH$_2$CHO) increase with increasing $T_{\rm max}$, while they in the second group (e.g., CH$_3$OH) decrease.
The temperature in the regions where they are effectively formed in the dust surfaces is relatively high (e.g., 25 - 40 K for NH$_2$CHO in Fig.~\ref{AbundancesOfCOMsInProtostellarCores}) in the first group, while it is relatively low (e.g., $\leq25$ K for CH$_3$OH in Fig.~\ref{AbundancesOfCOMsInProtostellarCores}) in the second group.
For COMs in the first group, the effective regions where COMs can be efficiently formed move outward, as $T_{\rm max}$ increases.
Since the infall velocity of the gas parcels is lower in the outer layers compared to in the inner layers, the gas parcels can stay for a longer time in the effective regions at high maximum temperature, producing more solid COMs.
Subsequently, the solid COMs are evaporated into the gas phase through thermal desorption reactions.
For the second group of COMs, the outer boundary of their effective regions ($\leq20$ K) where they are effectively formed on the dust surfaces beyond the outer boundary in this physical model (i.e., $r_{\rm out} = 1.686{\times}10^4$ AU).
Therefore, the effective regions gradually narrow with increasing $T_{\rm max}$, resulting in a short time for dust surface reactions and low abundances of COMs in ice mantles.
Due to the varying dependencies of peak abundances of different COMs on $T_{\rm max}$, the relation between the luminosity of the protostar and the abundance of COMs as well as the kind of species is difficult to find (e.g., \citealt{Higuchi+etal+2018,Yang+etal+2021}).
Overall, the variation of $T_{\rm max}$ cannot explain the chemical differences of WCCC sources and hot corino sources. 

\subsection{Dependence on Contraction Timescale}
\label{ContractionTimescale}
The contraction timescale ($t_{\rm cont}$) is an important physical parameter correlated to the protostellar mass to some extent (e.g., \citealt{Bernasconi+Maeder+1996}), so we investigated the influence of $t_{\rm cont}$ on WCCC and hot corino chemistry.
Figure~\ref{VariationOfContractionTime} shows the variation in peak abundances of carbon-chain molecules and COMs as a function of $t_{\rm cont}$ at some fixed values of $T_{\rm in}$.
At the larger $t_{\rm cont}$, it takes longer warm-up duration to reach the same temperature.
At the warm-up phase (e.g., $T_{\rm in} = 55$ K), the duration of the WCCC processes is prolonged with increasing $t_{\rm cont}$, so more carbon-chain molecules are produced.
However, if the duration is too long, the carbon-chain molecules gradually be consumed by some destruction reactions in the gas phase, as shown in the central temperature of 116 K.
At the final stage, the duration that the gas parcels stays in the WCCC regions where carbon-chain molecules are effectively generated is longer with a higher $t_{\rm cont}$, producing more abundant carbon-chain molecules.
A low $t_{\rm cont}$ for the high-mass protostar leads to scarce radical type carbon-chain molecules (e.g., CCH and CCS), which is consistent with some observations toward massive protostars (e.g., \citealt{Taniguchi+etal+2023}). 
At $t_{\rm cont} = 5{\times}10^5$ yr, the abundances of most carbon-chain molecules are higher than $10^{-9}$ at any moment.
The peak abundances of COMs vary with $t_{\rm cont}$ in a similar feature as carbon-chain molecules.
A long contraction timescale allows solid radical molecules to stay a long duration on the dust grains, leading to high peak abundances of COMs in ice mantles and gas phase.
Overall, the variation of $t_{\rm cont}$ can lead to chemical differences between WCCC sources and hot corino sources.

\section{Discussion}
\label{sect_Discussion}
\subsection{Explanation for the scarcity of COMs in prototypical WCCC sources}
\begin{figure} 
  \centering
  \includegraphics[width=\hsize, angle=0]{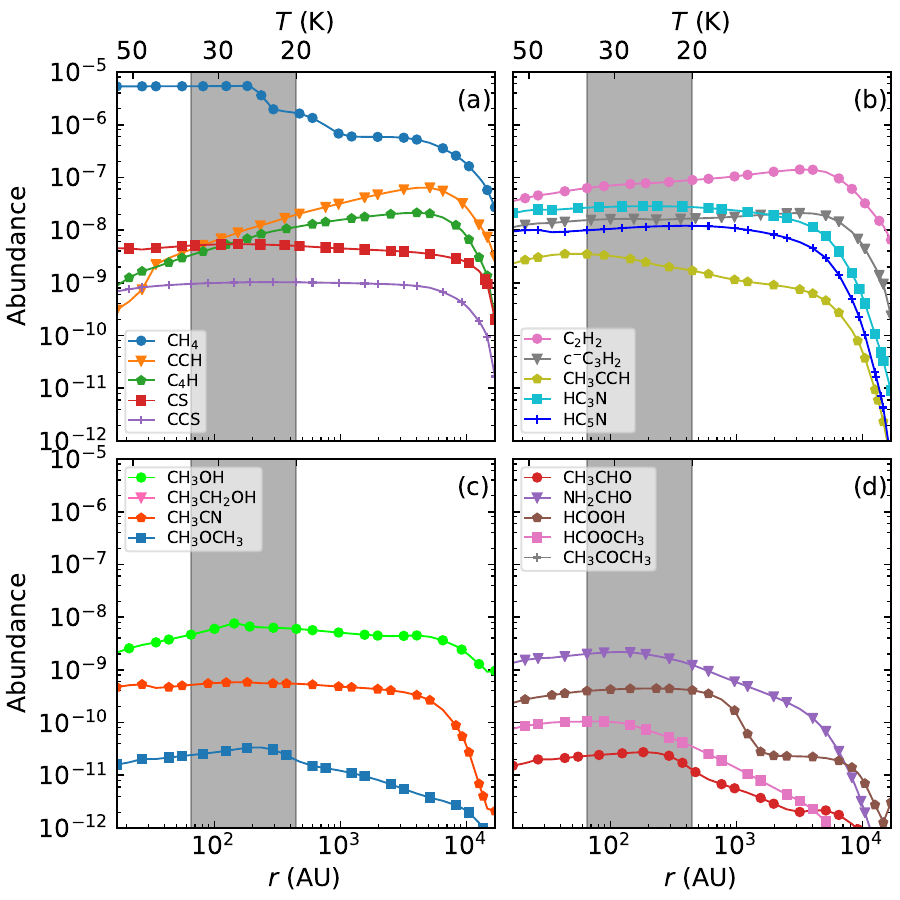}
  \caption{Radial distribution of gaseous carbon-chain molecules (a-b) and COMs (c-d) at $t = 2.7{\times}10^5$ yr in the case of $\zeta = 3{\times}10^{-16}$ s$^{-1}$ and $t_{\rm cont} = 5{\times}10^5$ yr. The dark gray shaded areas represent the WCCC regions ($T_{\rm dust} = 20 - 35$ K).} 
  \label{CCMAndCOMAtOneSpeTime}
\end{figure}
The relationship between the abundances of carbon-chain molecules and COMs is intricate, as demonstrated by studies conducted by \cite{Lindberg+etal+2016} and \cite{Higuchi+etal+2018}.
However, prototypical WCCC sources exhibit a clear chemical feature of abundant carbon-chain molecules and scarce COMs, such as L1527 and IRAS 15398, which is hard to reproduce this feature by chemical simulations (e.g., \citealt{Aikawa+etal+2020}).
Hence we attempt to answer this question.
The chemical structures are heavily depend on temperature profiles within a few hundred AU, as shown in Figs.~\ref{AbundancesOfCCMsInProtostellarCores} and \ref{AbundancesOfCOMsInProtostellarCores}.
When the dust temperature is lower than the evaporation temperature of specific COMs, their abundances can maintain at a low level (can see Fig.~\ref{AbundancesOfCOMsInProtostellarCores}).
If the dust temperature in the innermost regions exceeds $\sim30$ K but remains below $\sim100$ K, unsaturated carbon-chain molecules may be abundant while COMs are scarce, indicating the presence of prototypical WCCC characteristics.
This assumption may correspond to following two cases: the sublimation regions of COMs are apparently smaller than the beam size, or located within the rotationally supported disk (\citealt{Aikawa+etal+2020}). 
In the former scenario, the emissions of COMs can be difficult to detect due to the severe beam dilution under an insufficient spatial resolution. 
In the latter scenario, the spatial distribution of gaseous COMs is determined by the destruction timescale and the radial migration timescale of gas parcels within the disk.
If the migration timescale is much larger than the destruction timescale of COMs, the emissions of COMs in the disk are still difficult to detect due to low abundances.

Considering the influence of cosmic rays and the contraction timescale of the protostar on WCCC and hot corino chemistry in Sect.~\ref{DependenceOnSomePhysicalParameters}, we adopted a high cosmic-ray ionization rate of $3{\times}10^{-16}$ s$^{-1}$ and a long contraction timescale of $5{\times}10^5$ yr as an example to reproduce more abundant carbon-chain molecules and scarce COMs in prototypical WCCC sources.
Figure~\ref{CCMAndCOMAtOneSpeTime} shows the radial distribution of carbon-chain molecules and COMs at the moment of $2.7{\times}10^5$ yr in this case, when the innermost temperature is only $\sim 55$ K.
The ice CH$_4$ molecules are evaporated into the gas phase in WCCC regions through the thermal desorption mechanism, leading to a high abundance of gaseous CH$_4$ of $4{\times}10^{-6}$, which activates WCCC even in the vicinity of the protostar (i.e., $r \sim 100$ AU).
The abundances of long carbon-chain molecules are mostly above $10^{-9}$ in WCCC regions, while C$_n$H$_2$ and HC$_{2n+1}$N groups are also abundant in the compact regions.
Therefore, the protostar consists of a WCCC core, rich in small molecules, such as c-C$_3$H$_2$, and an extended envelope, rich in CCH and C$_4$H, which are consistent with observed features in IRAS 15398 and L1527 (e.g., \citealt{Sakai+etal+2009,Sakai+etal+2010,Araki+etal+2017,Higuchi+etal+2018}).
On the other hand, the abundances of most gaseous COMs at $r \leqslant 100$ AU regions are lower than $10^{-9}$ due to slow thermal migration and thermal desorption, so few COMs have been detected in prototypical WCCC sources as L1527 and IRAS 15398.
Overall, the scarcity of COMs in prototypical WCCC sources may be because that the temperature in the innermost protostellar envelopes has not reached lower limit required to motivate hot corino chemistry.
Meanwhile, this also suggests that the high cosmic ray ionization rate and the long contraction timescale of the protostar are favorable physical parameters responsible for the chemical characteristics in the prototypical WCCC sources.

\subsection{Differences between different gas-phase networks}
\label{InfluenceOfGasPhaseNetwork}
Only a small portion of the reactions included in currently available public networks (e.g., KIDA and UDFA) have been investigated either in the laboratory or theoretically, leading to a significant overall level of uncertainty.
\cite{Tinacci+etal+2023} used quantum mechanics calculations and found that only 5\% of the reactions in the KIDA2014 database, which were previously assumed to be barrierless, are actually endothermic.
After removing these endothermic reactions, they found that there was only a minor change to the abundance of most species, except for some Si-bearing species. 
As such, this treatment is not likely to significantly change our results. 
Recently, \cite{Millar+etal+2023} released the sixth version of the UDFA network, which includes more species and reactions compared to the last release. 
However, due to limited advancements in carbon-chain molecules, we continue to compare the UDFA12 with the KIDA2014 gas-phase network used in this work.

There are some differences between the UDFA and KIDA networks in several aspects. The KIDA network has a larger and more comprehensive network, which can be combined with some grain-surface reactions to study chemistry in planetary atmospheres (\citealt{Millar+etal+2023}). 
The KIDA network includes many unmeasured ion-neutral reactions, while the UDFA network relies more on detected molecules and well-studied reactions (\citealt{Wakelam+etal+2012,Millar+etal+2023}). 
Additionally, there are differences in approaches to estimating rate coefficients, as well as the choice of experimental rate coefficients (\citealt{Wakelam+etal+2012}).
These differences lead to variations in the carbon-chain chemistry in typical dark clouds as simulated by \cite{Wakelam+etal+2012} and \cite{McElroy+etal+2013}.

Considering the dominance of the WCCC mechanism in carbon-chain molecules in the protostellar cores, we analyzed the key reactions associated with WCCC to assess the impact of the differences between these two networks on our simulated results. 
The WCCC begins with gas-phase reactions of gaseous CH$_4$ molecules with C$^+$ ions, followed by a series of subsequent reactions. 
The reactions involving the precursor CH$_4$ molecules and C$^+$ ions in the two networks have some differences, leading to variations in the subsequent WCCC reactions. 
Both networks include identical reactions involving CH$_4$ and C$^+$ to produce C$_2$H$_2^+$ and C$_2$H$_3^+$, as well as crucial formation reactions of C$_2$H and C$_2$H$_2$ molecules.
However, there is a significant discrepancy in reaction~\ref{ReactionOfCnHWithH2} between these two networks. 
The rate coefficient for the reaction~\ref{ReactionOfCnHWithH2} associated with C$_2$H$_2$ in the KIDA network is expressed by $1.14{\times}10^{-11}{\rm exp}(-1300/T)$ cm$^3$ s$^{-1}$, whereas the rate coefficient in the UDFA network is calculated by $9.11{\times}10^{-13}(T/300)^{2.57}{\rm exp}(-130/T)$ cm$^3$ s$^{-1}$. 
For instance, at a temperature of 150 K, the former rate coefficient is $1.96{\times}10^{-15}$ cm$^3$ s$^{-1}$, whereas the latter rate coefficient is $6.45{\times}10^{-14}$ cm$^3$ s$^{-1}$.
On the other hand, the UDFA network does not include reaction~\ref{ReactionOfCnHWithH2} related to C$_n$H ($n>2$). 
Hence the dip structure of C$_n$H would not appear in models using the UDFA network. 
The UDFA network also lacks the crucial formation reactions~\ref{ReactionOfC2HwithHCNType} of HC$_{2n+1}$N ($n=2, 3, 4$) in hot environments. 
In general, one can expect some differences in abundances of carbon-chain molecules using the UDFA gas-phase network. 
However, the spatial structures of HC$_{2n+1}$N and C$_{2n}$H$_2$ would not undergo significant changes as these features are determined by the evaporation temperatures. 

The impact of several crucial physical parameters (i.e., $A_{\rm V}^{\rm amb}$, $\zeta$, $T_{\rm max}$, and $t_{\rm cont}$) on WCCC has been explored using the KIDA network in Sect.~\ref{DependenceOnSomePhysicalParameters}. 
Does the differences between the KIDA and UDFA network directly impact our understanding of WCCC? 
Ultraviolet photons have a primary influence on WCCC processes through the CO photodissociation reaction, as discussed in Sect.~\ref{VisualExtinction}. 
Since both networks have the same CO photodissociation rates, so their differences would not change the impact of UV photons on WCCC. 
Similarly, the direct and induced ionization reactions due to cosmic-rays for CO and CH$_4$ molecules are essentially the same in both networks. 
Therefore, one can expect similar conclusions about the influence of cosmic rays on WCCC using the UDFA network, as shown by \cite{Kalvans+2021}. 
The majority of gas-phase reactions governing WCCC are similar in both networks, so they may have comparable effects on WCCC when changing $T_{\rm max}$ and $t_{\rm cont}$ with different effective durations. 
Regarding hot corino chemistry, since most COMs are mainly dominated by surface reactions, the variations in these two gas-phase networks would not clearly impact our understanding of hot corino chemistry in this work.

\subsection{Uncertainties in physical models and evolutionary timescales}
Besides the treatment of the gas-phase network, there are some uncertainties in the physical models of core evolution and the timescales for various stages, which  have the potential to impact the simulated results of WCCC and hot corino chemistry.
Numerous works have studied the chemistry or the dynamics in prestellar cores using various physical models, including the Larson-Penston collapse model (\citealt{Aikawa+etal+2001}), Bonnor-Ebert sphere model (\citealt{Lee+etal+2004}), hydrostatic model (\citealt{Masunaga+Inutsuka+2000}), isothermal collapse model (\citealt{Keto+Caselli+2010}), free-fall collapse model (\citealt{Sun+Du+2022}), and magnetohydrodynamic model (\citealt{Ciolek+Basu+2000}). 
From the perspective of the chemistry, these physical models primarily differ in two important physical quantities: the number density and the visual extinction.
For example, in \cite{Sun+Du+2022}'s work, the average density of clouds is lower than that in our physical model, resulting in a slow chemical evolution.
However, due to significant depletion of most molecules in the late stages, as observed in \cite{Hirota+etal+2009}, there is no significant difference in the abundance of precursor CH$_4$ and radicals adsorbed onto dust grains. 
Hence this would not significantly change the subsequent characteristics of WCCC and hot corino chemistry in the protostellar core.
For instance, \cite{Kalvans+2021} used a single-point collapse model, and obtained similar effects of cosmic rays on WCCC.
Analogously, \cite{Aikawa+etal+2020} got similar role of the interstellar fields on WCCC and hot corino chemistry in protostellar cores, using the hydrostatic prestellar cores.
The impact of variations in protostellar core models has relatively minor effects on WCCC and hot corino chemistry, mainly due to two reasons.
The dynamics of inside-out collapse in the protostellar core, as simulated in several hydrodynamic (\citealt{Masunaga+Inutsuka+2000}) and even magnetohydrodynamic models (\citealt{Tomida+etal+2010}), bears resemblance to the expansion wave collapse model in this work.
For instance, the density profile near the center in \cite{Aikawa+etal+2008} is close to that we used.
Another important reason is that carbon-chain molecules and COMs in the protostellar envelope is more dependent on the dust temperature than the number density.

There are also uncertainties in the timescales of various evolutionary stages, including the prestellar phase, the isothermal collapse phase, the warm-up phase, and the hot corino phase.
Differences in the average density and dynamics of molecular clouds can lead to variations in the lifetimes of prestellar cores (\citealt{Ward-Thompson+etal+2007,Evans+etal+2009}). 
To understand the impact of variations in the prestellar timescale on the simulated results, we conducted additional models by adopting different retardation factors $B$ (1.0, 0.7, 0.5, and 0.3) for the prestellar core to determine their timescales $t_{\rm pre}$ (corresponding to 0.5, 0.7, 1.0, and 1.7 Myr, respectively).
Our models suggest that WCCC and hot corino chemistry are boosted as $t_{\rm pre}$ increases, which align with the simulated results reported by \cite{Aikawa+etal+2020}.
Due to the similar dynamics and the relatively short timescale during the isothermal collapse phase (\citealt{Larson+1969,Masunaga+Inutsuka+2000,Lee+etal+2004}), the impact of timescale variations on WCCC and hot corino chemistry can be considered negligible.
The differences in warm-up timescales can strongly affect WCCC and hot corino chemistry, as extensively studied in Sect.~\ref{ContractionTimescale}.
To some extent, this is also an important factor contributing to the chemical differences between low-mass and high-mass protostellar cores.
As for the effect of hot corino timescale, we can perform a simple analysis by referring to Figs.~\ref{AbundancesOfCCMsInProtostellarCores} and \ref{AbundancesOfCOMsInProtostellarCores}.
During the hot corino stage, the chemical characteristics of carbon-chain molecules and COMs gradually weaken over time, making it difficult to observe these molecules in the case of extended hot corino timescales.
On another note, this explains why all currently discovered WCCC and hot corino sources are located in Class 0/I phases.

\section{Conclusions}
\label{sect_Conclusions}
In this paper, we explored possible reasons for chemical differences between WCCC sources and hot corino sources, and explained the scarcity of COMs in the prototypical WCCC sources.
Our conclusions are as follows:
\begin{enumerate}
  \item Based on the spatial distribution of unsaturated carbon-chain molecules, they are divided into two types: extended distribution and a small central dip around the protostar (e.g., C$_n$H group), central condensation (e.g., HC$_{2n+1}$N, and C$_{2n}$H$_2$ groups).
  The appearance of the small central dip around the protostar for C$_n$H group is primarily dominated by the destruction reactions: C$_n$H + H$_2 \rightarrow$ C$_n$H$_2$ + H in the compact central regions. 
  Most COMs exhibit flat abundance profiles in inner regions, which is because infall timescales of gas parcels through hot corino regions are shorter than the destruction timescales of COMs.
  \item Our fiducial model calculated abundant carbon-chain molecules and COMs, suggesting that WCCC and hot corino chemistry can coexist in some protostars. Additionally, this model reproduced canonical characteristics of WCCC and hot corino chemistry in the hybrid source L483.
  \item Strong UV photons (i.e., in the case of low the visual extinction of ambient clouds $A_{\rm V}^{\rm amb}$) can rapidly dissociate CO molecules, producing abundant C$^+$ ions, which accelerates WCCC processes and enhances WCCC characteristics.
  At a high $A_{\rm V}^{\rm amb}$, scarce UV photons weaken the dissociation of COMs and their precursors, enhancing hot corino chemistry features.
  Therefore, UV photons heighten the WCCC features by accelerating the photodissociation of CO molecules, while they weaken the hot corino features.
  This explains the statistical characteristics of observations, where WCCC sources are often found at the cloud boundaries, while hot corino sources tend to be located inside dense clouds.
  \item Another physical parameter in local environments - the cosmic-ray ionization rate ($\zeta$) can also lead to differences in WCCC, but the response of hot corino chemistry to $\zeta$ variation is highly species-dependent.
  Intensive cosmic rays accelerate the dissociation of CO and CH$_4$ molecules, generating abundant C$^+$ ions, which enhances the WCCC characteristics.
  The influence of cosmic rays on hot corino chemistry is relatively complex, which is mainly because that COMs and their precursors are affected by various reactions associated with cosmic rays. 
  \item The maximum temperature at the warm-up phase ($T_{\rm max}$) cannot affect WCCC features, while it's influence varies for different COMs. 
  It is one reason for the weak relation between the luminosity of the protostar and the abundance of COMs, as well as the kind of species. 
  Overall, the variation of $T_{\rm max}$ cannot explain the chemical differences between WCCC sources and hot corino sources.
  The long contraction timescale ($t_{\rm cont}$) promotes WCCC and hot corino chemistry, by prolonging the effective duration of WCCC reactions in the gas phase and surface reactions of COMs, respectively.
  This suggests that the variation of $t_{\rm cont}$ can generate chemical differences between WCCC sources and hot corino sources to some extent.
  \item To explain the scarcity of COMs in prototypical WCCC sources (e.g., L1527, and IRAS 15398), we adopted $\zeta = 3{\times}10^{-16}$ s$^{-1}$ and $t_{\rm cont} = 5{\times}10^5$ yr.
  When the temperature in the innermost layers is only about 55 K, we reproduce the chemical characteristics (i.e., scarce COMs and abundant carbon-chain molecules) in prototypical WCCC sources.
  Hence, we suggest that scarce COMs in prototypical WCCC sources can be attributed to the insufficient dust temperature in the innermost envelopes to trigger hot corino chemistry.
  Furthermore, we propose that the high $\zeta$ and the long $t_{\rm cont}$ also are crucial physical parameters responsible to shape scarce COMs in the prototypical WCCC sources.
\end{enumerate}

\begin{acknowledgements}
  We are extremely grateful to the anonymous referee for critical and constructive comments that greatly improved the original manuscript.
  We thank Wasim Iqbal and Rong Ma for helpful discussions and valuable comments.
  F.D. acknowledges the support from the National Natural Science Foundation of China, grant No. 11873094 and 12041305.
  Y.W. acknowledges the support by the Natural Science Foundation
  of Jiangsu Province (Grant Number BK20221163).
  The Taurus High Performance Computing system of Xinjiang Astronomical Observatory was used in our simulations.
\end{acknowledgements}

%
%

\bibliographystyle{aa}
\bibliography{bibtex}

\begin{thebibliography}{148}
\expandafter\ifx\csname natexlab\endcsname\relax\def\natexlab#1{#1}\fi

\bibitem[{{Ag{\'u}ndez} {et~al.}(2015{\natexlab{a}}){Ag{\'u}ndez},
  {Cernicharo}, {de Vicente}, {Marcelino}, {Roueff}, {Fuente}, {Gerin},
  {Gu{\'e}lin}, {Albo}, {Barcia}, {Barbas}, {Bola{\~n}o}, {Colomer}, {Diez},
  {Gallego}, {G{\'o}mez-Gonz{\'a}lez}, {L{\'o}pez-Fern{\'a}ndez},
  {L{\'o}pez-Fern{\'a}ndez}, {L{\'o}pez-P{\'e}rez}, {Malo}, {Serna}, \&
  {Tercero}}]{Agundez+etal+2015b}
{Ag{\'u}ndez}, M., {Cernicharo}, J., {de Vicente}, P., {et~al.}
  2015{\natexlab{a}}, \aap, 579, L10

\bibitem[{{Ag{\'u}ndez} {et~al.}(2015{\natexlab{b}}){Ag{\'u}ndez},
  {Cernicharo}, \& {Gu{\'e}lin}}]{Agundez+etal+2015a}
{Ag{\'u}ndez}, M., {Cernicharo}, J., \& {Gu{\'e}lin}, M. 2015{\natexlab{b}},
  \aap, 577, L5

\bibitem[{{Ag{\'u}ndez} {et~al.}(2019){Ag{\'u}ndez}, {Marcelino}, {Cernicharo},
  {Roueff}, \& {Tafalla}}]{Agundez+etal+2019}
{Ag{\'u}ndez}, M., {Marcelino}, N., {Cernicharo}, J., {Roueff}, E., \&
  {Tafalla}, M. 2019, \aap, 625, A147

\bibitem[{{Aikawa} {et~al.}(2020){Aikawa}, {Furuya}, {Yamamoto}, \&
  {Sakai}}]{Aikawa+etal+2020}
{Aikawa}, Y., {Furuya}, K., {Yamamoto}, S., \& {Sakai}, N. 2020, \apj, 897, 110

\bibitem[{{Aikawa} {et~al.}(2001){Aikawa}, {Ohashi}, {Inutsuka}, {Herbst}, \&
  {Takakuwa}}]{Aikawa+etal+2001}
{Aikawa}, Y., {Ohashi}, N., {Inutsuka}, S.-i., {Herbst}, E., \& {Takakuwa}, S.
  2001, \apj, 552, 639

\bibitem[{{Aikawa} {et~al.}(2008){Aikawa}, {Wakelam}, {Garrod}, \&
  {Herbst}}]{Aikawa+etal+2008}
{Aikawa}, Y., {Wakelam}, V., {Garrod}, R.~T., \& {Herbst}, E. 2008, \apj, 674,
  984

\bibitem[{{Al-Halabi} \& {van Dishoeck}(2007)}]{Al-Halabi+vanDishoeck+2007}
{Al-Halabi}, A. \& {van Dishoeck}, E.~F. 2007, \mnras, 382, 1648

\bibitem[{{Allen} \& {Robinson}(1977)}]{Allen+Robinson+1977}
{Allen}, M. \& {Robinson}, G.~W. 1977, \apj, 212, 396

\bibitem[{{Andersson} \& {van Dishoeck}(2008)}]{Andersson+vanDishoeck+2008}
{Andersson}, S. \& {van Dishoeck}, E.~F. 2008, \aap, 491, 907

\bibitem[{{Araki} {et~al.}(2017){Araki}, {Takano}, {Sakai}, {Yamamoto},
  {Oyama}, {Kuze}, \& {Tsukiyama}}]{Araki+etal+2017}
{Araki}, M., {Takano}, S., {Sakai}, N., {et~al.} 2017, \apj, 847, 51

\bibitem[{{Barone} {et~al.}(2015){Barone}, {Latouche}, {Skouteris}, {Vazart},
  {Balucani}, {Ceccarelli}, \& {Lefloch}}]{Barone+etal+2015}
{Barone}, V., {Latouche}, C., {Skouteris}, D., {et~al.} 2015, \mnras, 453, L31

\bibitem[{{Belloche} {et~al.}(2020){Belloche}, {Maury}, {Maret}, {Anderl},
  {Bacmann}, {Andr{\'e}}, {Bontemps}, {Cabrit}, {Codella}, {Gaudel}, {Gueth},
  {Lef{\`e}vre}, {Lefloch}, {Podio}, \& {Testi}}]{Belloche+etal+2020}
{Belloche}, A., {Maury}, A.~J., {Maret}, S., {et~al.} 2020, \aap, 635, A198

\bibitem[{{Bergin} \& {Tafalla}(2007)}]{Bergin+Tafalla+2007}
{Bergin}, E.~A. \& {Tafalla}, M. 2007, \araa, 45, 339

\bibitem[{{Bernasconi} \& {Maeder}(1996)}]{Bernasconi+Maeder+1996}
{Bernasconi}, P.~A. \& {Maeder}, A. 1996, \aap, 307, 829

\bibitem[{{Bertin} {et~al.}(2013){Bertin}, {Fayolle}, {Romanzin}, {Poderoso},
  {Michaut}, {Philippe}, {Jeseck}, {{\"O}berg}, {Linnartz}, \&
  {Fillion}}]{Bertin+etal+2013}
{Bertin}, M., {Fayolle}, E.~C., {Romanzin}, C., {et~al.} 2013, \apj, 779, 120

\bibitem[{{Bianchi} {et~al.}(2019){Bianchi}, {Codella}, {Ceccarelli}, {Vazart},
  {Bachiller}, {Balucani}, {Bouvier}, {De Simone}, {Enrique-Romero}, {Kahane},
  {Lefloch}, {L{\'o}pez-Sepulcre}, {Ospina-Zamudio}, {Podio}, \&
  {Taquet}}]{Bianchi+etal+2019}
{Bianchi}, E., {Codella}, C., {Ceccarelli}, C., {et~al.} 2019, \mnras, 483,
  1850

\bibitem[{{Bohlin} {et~al.}(1978){Bohlin}, {Savage}, \&
  {Drake}}]{Bohlin+etal+1978}
{Bohlin}, R.~C., {Savage}, B.~D., \& {Drake}, J.~F. 1978, \apj, 224, 132

\bibitem[{{Bottinelli} {et~al.}(2004{\natexlab{a}}){Bottinelli}, {Ceccarelli},
  {Lefloch}, {Williams}, {Castets}, {Caux}, {Cazaux}, {Maret}, {Parise}, \&
  {Tielens}}]{Bottinelli+etal+2004a}
{Bottinelli}, S., {Ceccarelli}, C., {Lefloch}, B., {et~al.} 2004{\natexlab{a}},
  \apj, 615, 354

\bibitem[{{Bottinelli} {et~al.}(2004{\natexlab{b}}){Bottinelli}, {Ceccarelli},
  {Neri}, {Williams}, {Caux}, {Cazaux}, {Lefloch}, {Maret}, \&
  {Tielens}}]{Bottinelli+etal+2004b}
{Bottinelli}, S., {Ceccarelli}, C., {Neri}, R., {et~al.} 2004{\natexlab{b}},
  \apjl, 617, L69

\bibitem[{{Bottinelli} {et~al.}(2007){Bottinelli}, {Ceccarelli}, {Williams}, \&
  {Lefloch}}]{Bottinelli+etal+2007}
{Bottinelli}, S., {Ceccarelli}, C., {Williams}, J.~P., \& {Lefloch}, B. 2007,
  \aap, 463, 601

\bibitem[{{Bouvier} {et~al.}(2022){Bouvier}, {Ceccarelli},
  {L{\'o}pez-Sepulcre}, {Sakai}, {Yamamoto}, \& {Yang}}]{Bouvier+etal+2022}
{Bouvier}, M., {Ceccarelli}, C., {L{\'o}pez-Sepulcre}, A., {et~al.} 2022, \apj,
  929, 10

\bibitem[{{Bouvier} {et~al.}(2020){Bouvier}, {L{\'o}pez-Sepulcre},
  {Ceccarelli}, {Kahane}, {Imai}, {Sakai}, {Yamamoto}, \&
  {Dagdigian}}]{Bouvier+etal+2020}
{Bouvier}, M., {L{\'o}pez-Sepulcre}, A., {Ceccarelli}, C., {et~al.} 2020, \aap,
  636, A19

\bibitem[{{Caux} {et~al.}(2011){Caux}, {Kahane}, {Castets}, {Coutens},
  {Ceccarelli}, {Bacmann}, {Bisschop}, {Bottinelli}, {Comito}, {Helmich},
  {Lefloch}, {Parise}, {Schilke}, {Tielens}, {van Dishoeck}, {Vastel},
  {Wakelam}, \& {Walters}}]{Caux+etal+2011}
{Caux}, E., {Kahane}, C., {Castets}, A., {et~al.} 2011, \aap, 532, A23

\bibitem[{{Cazaux} {et~al.}(2003){Cazaux}, {Tielens}, {Ceccarelli}, {Castets},
  {Wakelam}, {Caux}, {Parise}, \& {Teyssier}}]{Cazaux+etal+2003}
{Cazaux}, S., {Tielens}, A.~G.~G.~M., {Ceccarelli}, C., {et~al.} 2003, \apjl,
  593, L51

\bibitem[{{Ceccarelli}(2004)}]{Ceccarelli+2004}
{Ceccarelli}, C. 2004, in Astronomical Society of the Pacific Conference
  Series, Vol. 323, Star Formation in the Interstellar Medium: In Honor of
  David Hollenbach, ed. D.~{Johnstone}, F.~C. {Adams}, D.~N.~C. {Lin}, D.~A.
  {Neufeeld}, \& E.~C. {Ostriker}, 195

\bibitem[{{Ceccarelli} {et~al.}(2017){Ceccarelli}, {Caselli}, {Fontani},
  {Neri}, {L{\'o}pez-Sepulcre}, {Codella}, {Feng}, {Jim{\'e}nez-Serra},
  {Lefloch}, {Pineda}, {Vastel}, {Alves}, {Bachiller}, {Balucani}, {Bianchi},
  {Bizzocchi}, {Bottinelli}, {Caux}, {Chac{\'o}n-Tanarro}, {Choudhury},
  {Coutens}, {Dulieu}, {Favre}, {Hily-Blant}, {Holdship}, {Kahane}, {Jaber
  Al-Edhari}, {Laas}, {Ospina}, {Oya}, {Podio}, {Pon}, {Punanova}, {Quenard},
  {Rimola}, {Sakai}, {Sims}, {Spezzano}, {Taquet}, {Testi}, {Theul{\'e}},
  {Ugliengo}, {Vasyunin}, {Viti}, {Wiesenfeld}, \&
  {Yamamoto}}]{Ceccarelli+etal+2017}
{Ceccarelli}, C., {Caselli}, P., {Fontani}, F., {et~al.} 2017, \apj, 850, 176

\bibitem[{{Ceccarelli} {et~al.}(2023){Ceccarelli}, {Codella}, {Balucani},
  {Bockelee-Morvan}, {Herbst}, {Vastel}, {Caselli}, {Favre}, {Lefloch},
  {Oberg}, \& {Yamamoto}}]{Ceccarelli+etal+2023}
{Ceccarelli}, C., {Codella}, C., {Balucani}, N., {et~al.} 2023, in Astronomical
  Society of the Pacific Conference Series, Vol. 534, Astronomical Society of
  the Pacific Conference Series, ed. S.~{Inutsuka}, Y.~{Aikawa}, T.~{Muto},
  K.~{Tomida}, \& M.~{Tamura}, 379

\bibitem[{{Chahine} {et~al.}(2022){Chahine}, {L{\'o}pez-Sepulcre}, {Neri},
  {Ceccarelli}, {Mercimek}, {Codella}, {Bouvier}, {Bianchi}, {Favre}, {Podio},
  {Alves}, {Sakai}, \& {Yamamoto}}]{Chahine+etal+2022}
{Chahine}, L., {L{\'o}pez-Sepulcre}, A., {Neri}, R., {et~al.} 2022, \aap, 657,
  A78

\bibitem[{{Chuang} {et~al.}(2018){Chuang}, {Fedoseev}, {Qasim}, {Ioppolo}, {van
  Dishoeck}, \& {Linnartz}}]{Chuang+etal+2018}
{Chuang}, K.~J., {Fedoseev}, G., {Qasim}, D., {et~al.} 2018, \apj, 853, 102

\bibitem[{{Ciolek} \& {Basu}(2000)}]{Ciolek+Basu+2000}
{Ciolek}, G.~E. \& {Basu}, S. 2000, \apj, 529, 925

\bibitem[{{Crimier} {et~al.}(2010){Crimier}, {Ceccarelli}, {Maret},
  {Bottinelli}, {Caux}, {Kahane}, {Lis}, \& {Olofsson}}]{Crimier+etal+2010}
{Crimier}, N., {Ceccarelli}, C., {Maret}, S., {et~al.} 2010, \aap, 519, A65

\bibitem[{{Dame} \& {Thaddeus}(1985)}]{Dame+Thaddeus+1985}
{Dame}, T.~M. \& {Thaddeus}, P. 1985, \apj, 297, 751

\bibitem[{{Draine}(1978)}]{Draine+1978}
{Draine}, B.~T. 1978, \apjs, 36, 595

\bibitem[{{Draine} \& {Bertoldi}(1996)}]{Draine+Bertoldi+1996}
{Draine}, B.~T. \& {Bertoldi}, F. 1996, \apj, 468, 269

\bibitem[{{Du}(2021)}]{Du+2021}
{Du}, F. 2021, Research in Astronomy and Astrophysics, 21, 077

\bibitem[{{Dupuy} {et~al.}(2017){Dupuy}, {Bertin}, {F{\'e}raud}, {Michaut},
  {Jeseck}, {Doronin}, {Philippe}, {Romanzin}, \& {Fillion}}]{Dupuy+etal+2017}
{Dupuy}, R., {Bertin}, M., {F{\'e}raud}, G., {et~al.} 2017, \aap, 603, A61

\bibitem[{{Evans} {et~al.}(2009){Evans}, {Dunham}, {J{\o}rgensen}, {Enoch},
  {Mer{\'\i}n}, {van Dishoeck}, {Alcal{\'a}}, {Myers}, {Stapelfeldt}, {Huard},
  {Allen}, {Harvey}, {van Kempen}, {Blake}, {Koerner}, {Mundy}, {Padgett}, \&
  {Sargent}}]{Evans+etal+2009}
{Evans}, Neal~J., I., {Dunham}, M.~M., {J{\o}rgensen}, J.~K., {et~al.} 2009,
  \apjs, 181, 321

\bibitem[{{Froebrich}(2005)}]{Froebrich+2005}
{Froebrich}, D. 2005, \apjs, 156, 169

\bibitem[{{Garrod}(2013)}]{Garrod+2013}
{Garrod}, R.~T. 2013, \apj, 765, 60

\bibitem[{{Garrod} \& {Herbst}(2006)}]{Garrod+Herbst+2006}
{Garrod}, R.~T. \& {Herbst}, E. 2006, \aap, 457, 927

\bibitem[{{Garrod} \& {Pauly}(2011)}]{Garrod+Pauly+2011}
{Garrod}, R.~T. \& {Pauly}, T. 2011, \apj, 735, 15

\bibitem[{{Garrod} {et~al.}(2007){Garrod}, {Wakelam}, \&
  {Herbst}}]{Garrod+etal+2007}
{Garrod}, R.~T., {Wakelam}, V., \& {Herbst}, E. 2007, \aap, 467, 1103

\bibitem[{{Garrod} {et~al.}(2008){Garrod}, {Widicus Weaver}, \&
  {Herbst}}]{Garrod+etal+2008}
{Garrod}, R.~T., {Widicus Weaver}, S.~L., \& {Herbst}, E. 2008, \apj, 682, 283

\bibitem[{{Graninger} {et~al.}(2016){Graninger}, {Wilkins}, \&
  {{\"O}berg}}]{Graninger+etal+2016}
{Graninger}, D.~M., {Wilkins}, O.~H., \& {{\"O}berg}, K.~I. 2016, \apj, 819,
  140

\bibitem[{{Habing}(1968)}]{Habing+1968}
{Habing}, H.~J. 1968, \bain, 19, 421

\bibitem[{{Hama} {et~al.}(2012){Hama}, {Kuwahata}, {Watanabe}, {Kouchi},
  {Kimura}, {Chigai}, \& {Pirronello}}]{Hama+etal+2012}
{Hama}, T., {Kuwahata}, K., {Watanabe}, N., {et~al.} 2012, \apj, 757, 185

\bibitem[{{Hasegawa} \& {Herbst}(1993)}]{Hasegawa+Herbst+1993}
{Hasegawa}, T.~I. \& {Herbst}, E. 1993, \mnras, 263, 589

\bibitem[{{Hasegawa} {et~al.}(1992){Hasegawa}, {Herbst}, \&
  {Leung}}]{Hasegawa+etal+1992}
{Hasegawa}, T.~I., {Herbst}, E., \& {Leung}, C.~M. 1992, \apjs, 82, 167

\bibitem[{{Hassel} {et~al.}(2008){Hassel}, {Herbst}, \&
  {Garrod}}]{Hassel+etal+2008}
{Hassel}, G.~E., {Herbst}, E., \& {Garrod}, R.~T. 2008, \apj, 681, 1385

\bibitem[{{He} {et~al.}(2018){He}, {Emtiaz}, \& {Vidali}}]{He+etal+2018}
{He}, J., {Emtiaz}, S., \& {Vidali}, G. 2018, \apj, 863, 156

\bibitem[{{He} {et~al.}(2015){He}, {Shi}, {Hopkins}, {Vidali}, \&
  {Kaufman}}]{He+etal+2015}
{He}, J., {Shi}, J., {Hopkins}, T., {Vidali}, G., \& {Kaufman}, M.~J. 2015,
  \apj, 801, 120

\bibitem[{{Herbst} \& {van Dishoeck}(2009)}]{Herbst+vanDishoeck+2009}
{Herbst}, E. \& {van Dishoeck}, E.~F. 2009, \araa, 47, 427

\bibitem[{{Higuchi} {et~al.}(2018){Higuchi}, {Sakai}, {Watanabe},
  {L{\'o}pez-Sepulcre}, {Yoshida}, {Oya}, {Imai}, {Zhang}, {Ceccarelli},
  {Lefloch}, {Codella}, {Bachiller}, {Hirota}, {Sakai}, \&
  {Yamamoto}}]{Higuchi+etal+2018}
{Higuchi}, A.~E., {Sakai}, N., {Watanabe}, Y., {et~al.} 2018, \apjs, 236, 52

\bibitem[{{Hirota} {et~al.}(2009){Hirota}, {Ohishi}, \&
  {Yamamoto}}]{Hirota+etal+2009}
{Hirota}, T., {Ohishi}, M., \& {Yamamoto}, S. 2009, \apj, 699, 585

\bibitem[{{Hocuk} {et~al.}(2017){Hocuk}, {Sz{\H{u}}cs}, {Caselli}, {Cazaux},
  {Spaans}, \& {Esplugues}}]{Hocuk+etal+2017}
{Hocuk}, S., {Sz{\H{u}}cs}, L., {Caselli}, P., {et~al.} 2017, \aap, 604, A58

\bibitem[{{Hollenbach} {et~al.}(1991){Hollenbach}, {Takahashi}, \&
  {Tielens}}]{Hollenbach+etal+1991}
{Hollenbach}, D.~J., {Takahashi}, T., \& {Tielens}, A.~G.~G.~M. 1991, \apj,
  377, 192

\bibitem[{{Imai} {et~al.}(2019){Imai}, {Oya}, {Sakai}, {L{\'o}pez-Sepulcre},
  {Watanabe}, \& {Yamamoto}}]{Imai+etal+2019}
{Imai}, M., {Oya}, Y., {Sakai}, N., {et~al.} 2019, \apjl, 873, L21

\bibitem[{{Imai} {et~al.}(2016){Imai}, {Sakai}, {Oya}, {L{\'o}pez-Sepulcre},
  {Watanabe}, {Ceccarelli}, {Lefloch}, {Caux}, {Vastel}, {Kahane}, {Sakai},
  {Hirota}, {Aikawa}, \& {Yamamoto}}]{Imai+etal+2016}
{Imai}, M., {Sakai}, N., {Oya}, Y., {et~al.} 2016, \apjl, 830, L37

\bibitem[{{Indriolo} {et~al.}(2010){Indriolo}, {Blake}, {Goto}, {Usuda}, {Oka},
  {Geballe}, {Fields}, \& {McCall}}]{Indriolo+etal+2010}
{Indriolo}, N., {Blake}, G.~A., {Goto}, M., {et~al.} 2010, \apj, 724, 1357

\bibitem[{{Indriolo} \& {McCall}(2012)}]{Indriolo+McCall+2012}
{Indriolo}, N. \& {McCall}, B.~J. 2012, \apj, 745, 91

\bibitem[{{Jacobsen} {et~al.}(2019){Jacobsen}, {J{\o}rgensen}, {Di Francesco},
  {Evans}, {Choi}, \& {Lee}}]{Jacobsen+etal+2019}
{Jacobsen}, S.~K., {J{\o}rgensen}, J.~K., {Di Francesco}, J., {et~al.} 2019,
  \aap, 629, A29

\bibitem[{{Jim{\'e}nez-Serra} {et~al.}(2016){Jim{\'e}nez-Serra}, {Vasyunin},
  {Caselli}, {Marcelino}, {Billot}, {Viti}, {Testi}, {Vastel}, {Lefloch}, \&
  {Bachiller}}]{Jimenez-Serra+etal+2016}
{Jim{\'e}nez-Serra}, I., {Vasyunin}, A.~I., {Caselli}, P., {et~al.} 2016,
  \apjl, 830, L6

\bibitem[{{J{\o}rgensen} {et~al.}(2020){J{\o}rgensen}, {Belloche}, \&
  {Garrod}}]{Jorgensen+etal+2020}
{J{\o}rgensen}, J.~K., {Belloche}, A., \& {Garrod}, R.~T. 2020, \araa, 58, 727

\bibitem[{{J{\o}rgensen} {et~al.}(2005){J{\o}rgensen}, {Bourke}, {Myers},
  {Sch{\"o}ier}, {van Dishoeck}, \& {Wilner}}]{Jorgensen+etal+2005}
{J{\o}rgensen}, J.~K., {Bourke}, T.~L., {Myers}, P.~C., {et~al.} 2005, \apj,
  632, 973

\bibitem[{{J{\o}rgensen} {et~al.}(2002){J{\o}rgensen}, {Sch{\"o}ier}, \& {van
  Dishoeck}}]{Jorgensen+etal+2002}
{J{\o}rgensen}, J.~K., {Sch{\"o}ier}, F.~L., \& {van Dishoeck}, E.~F. 2002,
  \aap, 389, 908

\bibitem[{{J{\o}rgensen} {et~al.}(2016){J{\o}rgensen}, {van der Wiel},
  {Coutens}, {Lykke}, {M{\"u}ller}, {van Dishoeck}, {Calcutt}, {Bjerkeli},
  {Bourke}, {Drozdovskaya}, {Favre}, {Fayolle}, {Garrod}, {Jacobsen},
  {{\"O}berg}, {Persson}, \& {Wampfler}}]{Jorgensen+etal+2016}
{J{\o}rgensen}, J.~K., {van der Wiel}, M.~H.~D., {Coutens}, A., {et~al.} 2016,
  \aap, 595, A117

\bibitem[{{J{\o}rgensen} {et~al.}(2013){J{\o}rgensen}, {Visser}, {Sakai},
  {Bergin}, {Brinch}, {Harsono}, {Lindberg}, {van Dishoeck}, {Yamamoto},
  {Bisschop}, \& {Persson}}]{Jorgensen+etal+2013}
{J{\o}rgensen}, J.~K., {Visser}, R., {Sakai}, N., {et~al.} 2013, \apjl, 779,
  L22

\bibitem[{{Kalv{\={a}}ns}(2018)}]{Kalvans+2018}
{Kalv{\={a}}ns}, J. 2018, \mnras, 478, 2753

\bibitem[{{Kalv{\={a}}ns}(2021)}]{Kalvans+2021}
{Kalv{\={a}}ns}, J. 2021, \apj, 910, 54

\bibitem[{{Kalv{\={a}}ns} {et~al.}(2017){Kalv{\={a}}ns}, {Shmeld}, {Kalnin}, \&
  {Hocuk}}]{Kalvans+etal+2017}
{Kalv{\={a}}ns}, J., {Shmeld}, I., {Kalnin}, J.~R., \& {Hocuk}, S. 2017,
  \mnras, 467, 1763

\bibitem[{{Karska} {et~al.}(2013){Karska}, {Herczeg}, {van Dishoeck},
  {Wampfler}, {Kristensen}, {Goicoechea}, {Visser}, {Nisini}, {San
  Jos{\'e}-Garc{\'\i}a}, {Bruderer}, {{\'S}niady}, {Doty}, {Fedele},
  {Y{\i}ld{\i}z}, {Benz}, {Bergin}, {Caselli}, {Herpin}, {Hogerheijde},
  {Johnstone}, {J{\o}rgensen}, {Liseau}, {Tafalla}, {van der Tak}, \&
  {Wyrowski}}]{Karska+etal+2013}
{Karska}, A., {Herczeg}, G.~J., {van Dishoeck}, E.~F., {et~al.} 2013, \aap,
  552, A141

\bibitem[{{Karssemeijer} \& {Cuppen}(2014)}]{Karssemeijer+Cuppen+2014}
{Karssemeijer}, L.~J. \& {Cuppen}, H.~M. 2014, \aap, 569, A107

\bibitem[{{Keto} \& {Caselli}(2010)}]{Keto+Caselli+2010}
{Keto}, E. \& {Caselli}, P. 2010, \mnras, 402, 1625

\bibitem[{{Kristensen} {et~al.}(2012){Kristensen}, {van Dishoeck}, {Bergin},
  {Visser}, {Y{\i}ld{\i}z}, {San Jose-Garcia}, {J{\o}rgensen}, {Herczeg},
  {Johnstone}, {Wampfler}, {Benz}, {Bruderer}, {Cabrit}, {Caselli}, {Doty},
  {Harsono}, {Herpin}, {Hogerheijde}, {Karska}, {van Kempen}, {Liseau},
  {Nisini}, {Tafalla}, {van der Tak}, \& {Wyrowski}}]{Kristensen+etal+2012}
{Kristensen}, L.~E., {van Dishoeck}, E.~F., {Bergin}, E.~A., {et~al.} 2012,
  \aap, 542, A8

\bibitem[{{Kristensen} {et~al.}(2010){Kristensen}, {van Dishoeck}, {van
  Kempen}, {Cuppen}, {Brinch}, {J{\o}rgensen}, \&
  {Hogerheijde}}]{Kristensen+etal+2010}
{Kristensen}, L.~E., {van Dishoeck}, E.~F., {van Kempen}, T.~A., {et~al.} 2010,
  \aap, 516, A57

\bibitem[{{Kuan} {et~al.}(2004){Kuan}, {Huang}, {Charnley}, {Hirano},
  {Takakuwa}, {Wilner}, {Liu}, {Ohashi}, {Bourke}, {Qi}, \&
  {Zhang}}]{Kuan+etal+2004}
{Kuan}, Y.-J., {Huang}, H.-C., {Charnley}, S.~B., {et~al.} 2004, \apjl, 616,
  L27

\bibitem[{{Larson}(1969)}]{Larson+1969}
{Larson}, R.~B. 1969, \mnras, 145, 271

\bibitem[{{Lee} {et~al.}(2019){Lee}, {Codella}, {Li}, \& {Liu}}]{Lee+etal+2019}
{Lee}, C.-F., {Codella}, C., {Li}, Z.-Y., \& {Liu}, S.-Y. 2019, \apj, 876, 63

\bibitem[{{Lee} {et~al.}(2004){Lee}, {Bergin}, \& {Evans}}]{Lee+etal+2004}
{Lee}, J.-E., {Bergin}, E.~A., \& {Evans}, Neal~J., I. 2004, \apj, 617, 360

\bibitem[{{Lefloch} {et~al.}(2018){Lefloch}, {Bachiller}, {Ceccarelli},
  {Cernicharo}, {Codella}, {Fuente}, {Kahane}, {L{\'o}pez-Sepulcre}, {Tafalla},
  {Vastel}, {Caux}, {Gonz{\'a}lez-Garc{\'\i}a}, {Bianchi}, {G{\'o}mez-Ruiz},
  {Holdship}, {Mendoza}, {Ospina-Zamudio}, {Podio}, {Qu{\'e}nard}, {Roueff},
  {Sakai}, {Viti}, {Yamamoto}, {Yoshida}, {Favre}, {Monfredini},
  {Quiti{\'a}n-Lara}, {Marcelino}, {Boechat-Roberty}, \&
  {Cabrit}}]{Lefloch+etal+2018}
{Lefloch}, B., {Bachiller}, R., {Ceccarelli}, C., {et~al.} 2018, \mnras, 477,
  4792

\bibitem[{{Li} {et~al.}(2013){Li}, {Heays}, {Visser}, {Ubachs}, {Lewis},
  {Gibson}, \& {van Dishoeck}}]{Li+etal+2013}
{Li}, X., {Heays}, A.~N., {Visser}, R., {et~al.} 2013, \aap, 555, A14

\bibitem[{{Lindberg} {et~al.}(2016){Lindberg}, {Charnley}, \&
  {Cordiner}}]{Lindberg+etal+2016}
{Lindberg}, J.~E., {Charnley}, S.~B., \& {Cordiner}, M.~A. 2016, \apjl, 833,
  L14

\bibitem[{{Lindberg} {et~al.}(2017){Lindberg}, {Charnley}, {J{\o}rgensen},
  {Cordiner}, \& {Bjerkeli}}]{Lindberg+etal+2017}
{Lindberg}, J.~E., {Charnley}, S.~B., {J{\o}rgensen}, J.~K., {Cordiner}, M.~A.,
  \& {Bjerkeli}, P. 2017, \apj, 835, 3

\bibitem[{{Loomis} {et~al.}(2018){Loomis}, {Cleeves}, {{\"O}berg}, {Aikawa},
  {Bergner}, {Furuya}, {Guzman}, \& {Walsh}}]{Loomis+etal+2018}
{Loomis}, R.~A., {Cleeves}, L.~I., {{\"O}berg}, K.~I., {et~al.} 2018, \apj,
  859, 131

\bibitem[{{L{\'o}pez-Sepulcre} {et~al.}(2017){L{\'o}pez-Sepulcre}, {Sakai},
  {Neri}, {Imai}, {Oya}, {Ceccarelli}, {Higuchi}, {Aikawa}, {Bottinelli},
  {Caux}, {Hirota}, {Kahane}, {Lefloch}, {Vastel}, {Watanabe}, \&
  {Yamamoto}}]{Lopez-Sepulcre+etal+2017}
{L{\'o}pez-Sepulcre}, A., {Sakai}, N., {Neri}, R., {et~al.} 2017, \aap, 606,
  A121

\bibitem[{{Marcelino} {et~al.}(2018{\natexlab{a}}){Marcelino}, {Ag{\'u}ndez},
  {Cernicharo}, {Roueff}, \& {Tafalla}}]{Marcelino+etal+2018b}
{Marcelino}, N., {Ag{\'u}ndez}, M., {Cernicharo}, J., {Roueff}, E., \&
  {Tafalla}, M. 2018{\natexlab{a}}, \aap, 612, L10

\bibitem[{{Marcelino} {et~al.}(2018{\natexlab{b}}){Marcelino}, {Gerin},
  {Cernicharo}, {Fuente}, {Wootten}, {Chapillon}, {Pety}, {Lis}, {Roueff},
  {Commer{\c{c}}on}, \& {Ciardi}}]{Marcelino+etal+2018a}
{Marcelino}, N., {Gerin}, M., {Cernicharo}, J., {et~al.} 2018{\natexlab{b}},
  \aap, 620, A80

\bibitem[{{Maret} {et~al.}(2005){Maret}, {Ceccarelli}, {Tielens}, {Caux},
  {Lefloch}, {Faure}, {Castets}, \& {Flower}}]{Maret+etal+2005}
{Maret}, S., {Ceccarelli}, C., {Tielens}, A.~G.~G.~M., {et~al.} 2005, \aap,
  442, 527

\bibitem[{{Mart{\'\i}n-Dom{\'e}nech} {et~al.}(2015){Mart{\'\i}n-Dom{\'e}nech},
  {Manzano-Santamar{\'\i}a}, {Mu{\~n}oz Caro}, {Cruz-D{\'\i}az}, {Chen},
  {Herrero}, \& {Tanarro}}]{Martin-Domenech+etal+2015}
{Mart{\'\i}n-Dom{\'e}nech}, R., {Manzano-Santamar{\'\i}a}, J., {Mu{\~n}oz
  Caro}, G.~M., {et~al.} 2015, \aap, 584, A14

\bibitem[{{Masunaga} \& {Inutsuka}(2000)}]{Masunaga+Inutsuka+2000}
{Masunaga}, H. \& {Inutsuka}, S.-i. 2000, \apj, 531, 350

\bibitem[{{McElroy} {et~al.}(2013){McElroy}, {Walsh}, {Markwick}, {Cordiner},
  {Smith}, \& {Millar}}]{McElroy+etal+2013}
{McElroy}, D., {Walsh}, C., {Markwick}, A.~J., {et~al.} 2013, \aap, 550, A36

\bibitem[{{Millar} {et~al.}(2023){Millar}, {Walsh}, {Van de Sande}, \&
  {Markwick}}]{Millar+etal+2023}
{Millar}, T.~J., {Walsh}, C., {Van de Sande}, M., \& {Markwick}, A.~J. 2023,
  arXiv e-prints, arXiv:2311.03936

\bibitem[{{Minissale} {et~al.}(2016{\natexlab{a}}){Minissale}, {Congiu}, \&
  {Dulieu}}]{Minissale+etal+2016a}
{Minissale}, M., {Congiu}, E., \& {Dulieu}, F. 2016{\natexlab{a}}, \aap, 585,
  A146

\bibitem[{{Minissale} {et~al.}(2016{\natexlab{b}}){Minissale}, {Dulieu},
  {Cazaux}, \& {Hocuk}}]{Minissale+etal+2016b}
{Minissale}, M., {Dulieu}, F., {Cazaux}, S., \& {Hocuk}, S. 2016{\natexlab{b}},
  \aap, 585, A24

\bibitem[{{Molinari} {et~al.}(2000){Molinari}, {Brand}, {Cesaroni}, \&
  {Palla}}]{Molinari+etal+2000}
{Molinari}, S., {Brand}, J., {Cesaroni}, R., \& {Palla}, F. 2000, \aap, 355,
  617

\bibitem[{{Murillo} {et~al.}(2018){Murillo}, {van Dishoeck}, {van der Wiel},
  {J{\o}rgensen}, {Drozdovskaya}, {Calcutt}, \& {Harsono}}]{Murillo+etal+2018}
{Murillo}, N.~M., {van Dishoeck}, E.~F., {van der Wiel}, M.~H.~D., {et~al.}
  2018, \aap, 617, A120

\bibitem[{{Myers} {et~al.}(1995){Myers}, {Bachiller}, {Caselli}, {Fuller},
  {Mardones}, {Tafalla}, \& {Wilner}}]{Myers+etal+1995}
{Myers}, P.~C., {Bachiller}, R., {Caselli}, P., {et~al.} 1995, \apjl, 449, L65

\bibitem[{{Nazari} {et~al.}(2021){Nazari}, {van Gelder}, {van Dishoeck},
  {Tabone}, {van't Hoff}, {Ligterink}, {Beuther}, {Boogert}, {Caratti o
  Garatti}, {Klaassen}, {Linnartz}, {Taquet}, \&
  {Tychoniec}}]{Nazari+etal+2021}
{Nazari}, P., {van Gelder}, M.~L., {van Dishoeck}, E.~F., {et~al.} 2021, \aap,
  650, A150

\bibitem[{{Neufeld} {et~al.}(2010){Neufeld}, {Goicoechea}, {Sonnentrucker},
  {Black}, {Pearson}, {Yu}, {Phillips}, {Lis}, {de Luca}, {Herbst}, {Rimmer},
  {Gerin}, {Bell}, {Boulanger}, {Cernicharo}, {Coutens}, {Dartois},
  {Kazmierczak}, {Encrenaz}, {Falgarone}, {Geballe}, {Giesen}, {Godard},
  {Goldsmith}, {Gry}, {Gupta}, {Hennebelle}, {Hily-Blant}, {Joblin},
  {Ko{\l}os}, {Kre{\l}owski}, {Mart{\'\i}n-Pintado}, {Menten}, {Monje},
  {Mookerjea}, {Perault}, {Persson}, {Plume}, {Salez}, {Schlemmer}, {Schmidt},
  {Stutzki}, {Teyssier}, {Vastel}, {Cros}, {Klein}, {Lorenzani}, {Philipp},
  {Samoska}, {Shipman}, {Tielens}, {Szczerba}, \&
  {Zmuidzinas}}]{Neufeld+etal+2010}
{Neufeld}, D.~A., {Goicoechea}, J.~R., {Sonnentrucker}, P., {et~al.} 2010,
  \aap, 521, L10

\bibitem[{{Neufeld} \& {Wolfire}(2017)}]{Neufeld+Wolfire+2017}
{Neufeld}, D.~A. \& {Wolfire}, M.~G. 2017, \apj, 845, 163

\bibitem[{{{\"O}berg} \& {Bergin}(2021)}]{Oberg+Bergin+2021}
{{\"O}berg}, K.~I. \& {Bergin}, E.~A. 2021, \physrep, 893, 1

\bibitem[{{{\"O}berg} {et~al.}(2008){{\"O}berg}, {Boogert}, {Pontoppidan},
  {Blake}, {Evans}, {Lahuis}, \& {van Dishoeck}}]{Oberg+etal+2008}
{{\"O}berg}, K.~I., {Boogert}, A.~C.~A., {Pontoppidan}, K.~M., {et~al.} 2008,
  \apj, 678, 1032

\bibitem[{{{\"O}berg} {et~al.}(2009{\natexlab{a}}){{\"O}berg}, {Garrod}, {van
  Dishoeck}, \& {Linnartz}}]{Oberg+etal+2009a}
{{\"O}berg}, K.~I., {Garrod}, R.~T., {van Dishoeck}, E.~F., \& {Linnartz}, H.
  2009{\natexlab{a}}, \aap, 504, 891

\bibitem[{{{\"O}berg} {et~al.}(2009{\natexlab{b}}){{\"O}berg}, {Linnartz},
  {Visser}, \& {van Dishoeck}}]{Oberg+etal+2009c}
{{\"O}berg}, K.~I., {Linnartz}, H., {Visser}, R., \& {van Dishoeck}, E.~F.
  2009{\natexlab{b}}, \apj, 693, 1209

\bibitem[{{{\"O}berg} {et~al.}(2009{\natexlab{c}}){{\"O}berg}, {van Dishoeck},
  \& {Linnartz}}]{Oberg+etal+2009b}
{{\"O}berg}, K.~I., {van Dishoeck}, E.~F., \& {Linnartz}, H.
  2009{\natexlab{c}}, \aap, 496, 281

\bibitem[{{Ohashi} {et~al.}(1997){Ohashi}, {Hayashi}, {Ho}, \&
  {Momose}}]{Ohashi+etal+1997}
{Ohashi}, N., {Hayashi}, M., {Ho}, P. T.~P., \& {Momose}, M. 1997, \apj, 475,
  211

\bibitem[{{Oya} {et~al.}(2017){Oya}, {Sakai}, {Watanabe}, {Higuchi}, {Hirota},
  {L{\'o}pez-Sepulcre}, {Sakai}, {Aikawa}, {Ceccarelli}, {Lefloch}, {Caux},
  {Vastel}, {Kahane}, \& {Yamamoto}}]{Oya+etal+2017}
{Oya}, Y., {Sakai}, N., {Watanabe}, Y., {et~al.} 2017, \apj, 837, 174

\bibitem[{{Padovani} {et~al.}(2018){Padovani}, {Ivlev}, {Galli}, \&
  {Caselli}}]{Padovani+etal+2018}
{Padovani}, M., {Ivlev}, A.~V., {Galli}, D., \& {Caselli}, P. 2018, \aap, 614,
  A111

\bibitem[{{Penston}(1969)}]{Penston+1969}
{Penston}, M.~V. 1969, \mnras, 144, 425

\bibitem[{{Pineda} {et~al.}(2012){Pineda}, {Maury}, {Fuller}, {Testi},
  {Garc{\'\i}a-Appadoo}, {Peck}, {Villard}, {Corder}, {van Kempen}, {Turner},
  {Tachihara}, \& {Dent}}]{Pineda+etal+2012}
{Pineda}, J.~E., {Maury}, A.~J., {Fuller}, G.~A., {et~al.} 2012, \aap, 544, L7

\bibitem[{{Rawlings} {et~al.}(1992){Rawlings}, {Hartquist}, {Menten}, \&
  {Williams}}]{Rawlings+etal+1992}
{Rawlings}, J.~M.~C., {Hartquist}, T.~W., {Menten}, K.~M., \& {Williams}, D.~A.
  1992, \mnras, 255, 471

\bibitem[{{Rowan-Robinson}(1980)}]{Rowan-Robinson+1980}
{Rowan-Robinson}, M. 1980, \apjs, 44, 403

\bibitem[{{Ruaud} {et~al.}(2015){Ruaud}, {Loison}, {Hickson}, {Gratier},
  {Hersant}, \& {Wakelam}}]{Ruaud+etal+2015}
{Ruaud}, M., {Loison}, J.~C., {Hickson}, K.~M., {et~al.} 2015, \mnras, 447,
  4004

\bibitem[{{Ruaud} {et~al.}(2016){Ruaud}, {Wakelam}, \&
  {Hersant}}]{Ruaud+etal+2016}
{Ruaud}, M., {Wakelam}, V., \& {Hersant}, F. 2016, \mnras, 459, 3756

\bibitem[{{Ruffle} \& {Herbst}(2000)}]{Ruffle+Herbst+2000}
{Ruffle}, D.~P. \& {Herbst}, E. 2000, \mnras, 319, 837

\bibitem[{{Sakai} {et~al.}(2009){Sakai}, {Sakai}, {Hirota}, {Burton}, \&
  {Yamamoto}}]{Sakai+etal+2009}
{Sakai}, N., {Sakai}, T., {Hirota}, T., {Burton}, M., \& {Yamamoto}, S. 2009,
  \apj, 697, 769

\bibitem[{{Sakai} {et~al.}(2008){Sakai}, {Sakai}, {Hirota}, \&
  {Yamamoto}}]{Sakai+etal+2008}
{Sakai}, N., {Sakai}, T., {Hirota}, T., \& {Yamamoto}, S. 2008, \apj, 672, 371

\bibitem[{{Sakai} {et~al.}(2010){Sakai}, {Sakai}, {Hirota}, \&
  {Yamamoto}}]{Sakai+etal+2010}
{Sakai}, N., {Sakai}, T., {Hirota}, T., \& {Yamamoto}, S. 2010, \apj, 722, 1633

\bibitem[{{Sakai} {et~al.}(2006){Sakai}, {Sakai}, \&
  {Yamamoto}}]{Sakai+etal+2006}
{Sakai}, N., {Sakai}, T., \& {Yamamoto}, S. 2006, \pasj, 58, L15

\bibitem[{{Sakai} \& {Yamamoto}(2013)}]{Sakai+Yamamoto+2013}
{Sakai}, N. \& {Yamamoto}, S. 2013, Chemical Reviews, 113, 8981

\bibitem[{{Scibelli} {et~al.}(2021){Scibelli}, {Shirley}, {Vasyunin}, \&
  {Launhardt}}]{Scibelli+etal+2021}
{Scibelli}, S., {Shirley}, Y., {Vasyunin}, A., \& {Launhardt}, R. 2021, \mnras,
  504, 5754

\bibitem[{{Shen} {et~al.}(2004){Shen}, {Greenberg}, {Schutte}, \& {van
  Dishoeck}}]{Shen+etal+2004}
{Shen}, C.~J., {Greenberg}, J.~M., {Schutte}, W.~A., \& {van Dishoeck}, E.~F.
  2004, \aap, 415, 203

\bibitem[{{Shu}(1977)}]{Shu+1977}
{Shu}, F.~H. 1977, \apj, 214, 488

\bibitem[{{Shu} {et~al.}(1987){Shu}, {Adams}, \& {Lizano}}]{Shu+etal+1987}
{Shu}, F.~H., {Adams}, F.~C., \& {Lizano}, S. 1987, \araa, 25, 23

\bibitem[{{Sicilia-Aguilar} {et~al.}(2019){Sicilia-Aguilar}, {Patel}, {Fang},
  {Roccatagliata}, {Getman}, \& {Goldsmith}}]{Sicilia-Aguilar+etal+2019}
{Sicilia-Aguilar}, A., {Patel}, N., {Fang}, M., {et~al.} 2019, \aap, 622, A118

\bibitem[{{Spezzano} {et~al.}(2016){Spezzano}, {Bizzocchi}, {Caselli}, {Harju},
  \& {Br{\"u}nken}}]{Spezzano+etal+2016}
{Spezzano}, S., {Bizzocchi}, L., {Caselli}, P., {Harju}, J., \& {Br{\"u}nken},
  S. 2016, \aap, 592, L11

\bibitem[{{Spitzer}(1978)}]{Spitzer+1978}
{Spitzer}, L. 1978, {Physical processes in the interstellar medium} (Wiley)

\bibitem[{{Sun} \& {Du}(2022)}]{Sun+Du+2022}
{Sun}, J. \& {Du}, F. 2022, Research in Astronomy and Astrophysics, 22, 065022

\bibitem[{{Taniguchi} {et~al.}(2019){Taniguchi}, {Herbst}, {Caselli},
  {Paulive}, {Maffucci}, \& {Saito}}]{Taniguchi+etal+2019}
{Taniguchi}, K., {Herbst}, E., {Caselli}, P., {et~al.} 2019, \apj, 881, 57

\bibitem[{{Taniguchi} {et~al.}(2023){Taniguchi}, {Majumdar}, {Caselli},
  {Takakuwa}, {Hsieh}, {Saito}, {Li}, {Dobashi}, {Shimoikura}, {Nakamura},
  {Tan}, \& {Herbst}}]{Taniguchi+etal+2023}
{Taniguchi}, K., {Majumdar}, L., {Caselli}, P., {et~al.} 2023, \apjs, 267, 4

\bibitem[{{Taniguchi} {et~al.}(2021{\natexlab{a}}){Taniguchi}, {Majumdar},
  {Plunkett}, {Takakuwa}, {Lis}, {Goldsmith}, {Nakamura}, {Saito}, \&
  {Herbst}}]{Taniguchi+etal+2021b}
{Taniguchi}, K., {Majumdar}, L., {Plunkett}, A., {et~al.} 2021{\natexlab{a}},
  \apj, 922, 152

\bibitem[{{Taniguchi} {et~al.}(2021{\natexlab{b}}){Taniguchi}, {Majumdar},
  {Takakuwa}, {Saito}, {Lis}, {Goldsmith}, \& {Herbst}}]{Taniguchi+etal+2021a}
{Taniguchi}, K., {Majumdar}, L., {Takakuwa}, S., {et~al.} 2021{\natexlab{b}},
  \apj, 910, 141

\bibitem[{{Taniguchi} {et~al.}(2018){Taniguchi}, {Saito}, {Sridharan}, \&
  {Minamidani}}]{Taniguchi+etal+2018}
{Taniguchi}, K., {Saito}, M., {Sridharan}, T.~K., \& {Minamidani}, T. 2018,
  \apj, 854, 133

\bibitem[{{Taquet} {et~al.}(2015){Taquet}, {L{\'o}pez-Sepulcre}, {Ceccarelli},
  {Neri}, {Kahane}, \& {Charnley}}]{Taquet+etal+2015}
{Taquet}, V., {L{\'o}pez-Sepulcre}, A., {Ceccarelli}, C., {et~al.} 2015, \apj,
  804, 81

\bibitem[{{Tielens} \& {Hagen}(1982)}]{Tielens+Hagen+1982}
{Tielens}, A.~G.~G.~M. \& {Hagen}, W. 1982, \aap, 114, 245

\bibitem[{{Tinacci} {et~al.}(2023){Tinacci}, {Ferrada-Chamorro}, {Ceccarelli},
  {Pantaleone}, {Ascenzi}, {Maranzana}, {Balucani}, \&
  {Ugliengo}}]{Tinacci+etal+2023}
{Tinacci}, L., {Ferrada-Chamorro}, S., {Ceccarelli}, C., {et~al.} 2023, \apjs,
  266, 38

\bibitem[{{Tomida} {et~al.}(2010){Tomida}, {Tomisaka}, {Matsumoto}, {Ohsuga},
  {Machida}, \& {Saigo}}]{Tomida+etal+2010}
{Tomida}, K., {Tomisaka}, K., {Matsumoto}, T., {et~al.} 2010, \apjl, 714, L58

\bibitem[{{van Dishoeck} {et~al.}(1995){van Dishoeck}, {Blake}, {Jansen}, \&
  {Groesbeck}}]{vanDishoeck+etal+1995}
{van Dishoeck}, E.~F., {Blake}, G.~A., {Jansen}, D.~J., \& {Groesbeck}, T.~D.
  1995, \apj, 447, 760

\bibitem[{{Visser} {et~al.}(2009){Visser}, {van Dishoeck}, \&
  {Black}}]{Visser+etal+2009}
{Visser}, R., {van Dishoeck}, E.~F., \& {Black}, J.~H. 2009, \aap, 503, 323

\bibitem[{{Viti} {et~al.}(2004){Viti}, {Collings}, {Dever}, {McCoustra}, \&
  {Williams}}]{Viti+etal+2004}
{Viti}, S., {Collings}, M.~P., {Dever}, J.~W., {McCoustra}, M. R.~S., \&
  {Williams}, D.~A. 2004, \mnras, 354, 1141

\bibitem[{{Viti} \& {Williams}(1999)}]{Viti+Williams+1999}
{Viti}, S. \& {Williams}, D.~A. 1999, \mnras, 305, 755

\bibitem[{{Wakelam} {et~al.}(2012){Wakelam}, {Herbst}, {Loison}, {Smith},
  {Chandrasekaran}, {Pavone}, {Adams}, {Bacchus-Montabonel}, {Bergeat},
  {B{\'e}roff}, {Bierbaum}, {Chabot}, {Dalgarno}, {van Dishoeck}, {Faure},
  {Geppert}, {Gerlich}, {Galli}, {H{\'e}brard}, {Hersant}, {Hickson},
  {Honvault}, {Klippenstein}, {Le Picard}, {Nyman}, {Pernot}, {Schlemmer},
  {Selsis}, {Sims}, {Talbi}, {Tennyson}, {Troe}, {Wester}, \&
  {Wiesenfeld}}]{Wakelam+etal+2012}
{Wakelam}, V., {Herbst}, E., {Loison}, J.~C., {et~al.} 2012, \apjs, 199, 21

\bibitem[{{Wakelam} {et~al.}(2015){Wakelam}, {Loison}, {Herbst}, {Pavone},
  {Bergeat}, {B{\'e}roff}, {Chabot}, {Faure}, {Galli}, {Geppert}, {Gerlich},
  {Gratier}, {Harada}, {Hickson}, {Honvault}, {Klippenstein}, {Le Picard},
  {Nyman}, {Ruaud}, {Schlemmer}, {Sims}, {Talbi}, {Tennyson}, \&
  {Wester}}]{Wakelam+etal+2015}
{Wakelam}, V., {Loison}, J.~C., {Herbst}, E., {et~al.} 2015, \apjs, 217, 20

\bibitem[{{Wang} {et~al.}(2019){Wang}, {Chang}, \& {Wang}}]{Wang+etal+2019}
{Wang}, Y., {Chang}, Q., \& {Wang}, H. 2019, \aap, 622, A185

\bibitem[{{Ward-Thompson} {et~al.}(2007){Ward-Thompson}, {Andr{\'e}},
  {Crutcher}, {Johnstone}, {Onishi}, \& {Wilson}}]{Ward-Thompson+etal+2007}
{Ward-Thompson}, D., {Andr{\'e}}, P., {Crutcher}, R., {et~al.} 2007, in
  Protostars and Planets V, ed. B.~{Reipurth}, D.~{Jewitt}, \& K.~{Keil}, 33

\bibitem[{{Watanabe} {et~al.}(2003){Watanabe}, {Shiraki}, \&
  {Kouchi}}]{Watanabe+etal+2003}
{Watanabe}, N., {Shiraki}, T., \& {Kouchi}, A. 2003, \apjl, 588, L121

\bibitem[{{Yang} {et~al.}(2021){Yang}, {Sakai}, {Zhang}, {Murillo}, {Zhang},
  {Higuchi}, {Zeng}, {L{\'o}pez-Sepulcre}, {Yamamoto}, {Lefloch}, {Bouvier},
  {Ceccarelli}, {Hirota}, {Imai}, {Oya}, {Sakai}, \&
  {Watanabe}}]{Yang+etal+2021}
{Yang}, Y.-L., {Sakai}, N., {Zhang}, Y., {et~al.} 2021, \apj, 910, 20

\bibitem[{{Zucconi} {et~al.}(2001){Zucconi}, {Walmsley}, \&
  {Galli}}]{Zucconi+etal+2001}
{Zucconi}, A., {Walmsley}, C.~M., \& {Galli}, D. 2001, \aap, 376, 650

\end{thebibliography}

\end{document}